\pdfoutput=1
\documentclass[12pt,tightenlines,longbibliography,amsmath,amsthm,amssymb,eqsecnum,showkeys,secnumarabic,notitlepage,superscriptaddress]{revtex4-1}

\usepackage{hyperref}
\let\cites=\cite
\let\address=\affiliation

\newcommand{\up}{\uparrow}

\DeclareMathOperator{\Tr}{Tr}
\DeclareMathOperator{\sign}{sign}

\makeatletter
\renewcommand{\p@subsection}{}
\renewcommand{\p@subsubsection}{}
\makeatother

\begin{document}

\title{Time and temperature-dependent correlation function of an impurity in a one-dimensional Fermi gas as a Fredholm determinant}

\author{Oleksandr Gamayun}
\address{Instituut-Lorentz, Universiteit Leiden, P.O. Box 9506, 2300 RA Leiden, The Netherlands}
\email{Gamayun@Lorentz.LeidenUniv.nl}

\author{Andrei G. Pronko}
\address{Steklov Mathematical Institute,
Fontanka 27, St.~Petersburg, 191023, Russia}
\email{a.g.pronko@gmail.com}

\author{Mikhail B. Zvonarev}
\address{LPTMS, CNRS, Univ. Paris-Sud, Universit\'e Paris-Saclay, 91405 Orsay, France}
\address{ITMO University, 197101, Saint-Petersburg, Russia}
\email{mikhail.zvonarev@gmail.com}

\begin{abstract}
We investigate a free one-dimensional spinless Fermi gas, and the Tonks-Girardeau gas interacting with a single impurity particle of equal mass. We obtain a Fredholm determinant representation for the time-dependent correlation function of the impurity particle. This representation is valid for an arbitrary temperature and an arbitrary repulsive or attractive impurity-gas $\delta$-function interaction potential. It includes, as particular cases, the representations obtained for zero temperature and arbitrary repulsion in [Nucl.\ Phys.\ B \textbf{892}, 83 (2015)], and for arbitrary temperature and infinite repulsion in [Nucl.\ Phys.\ B \textbf{520}, 594 (1998)].
\end{abstract}


\maketitle
\tableofcontents

\section{Introduction}

Quantum many-body interacting systems in one spatial dimension solved by the Bethe Ansatz are unique in that their complete set of eigenfunctions and eigenstates are known explicitly~\cite{korepin_book}. The use of this property constitutes a whole research field by now. Describing the time evolution of the observables is one of the most challenging problems in the field, remaining largely open.

The present paper is devoted to a particular Bethe Ansatz-solvable model: a single mobile impurity interacting with a free Fermi gas in one spatial dimension through a $\delta$-function potential of an arbitrary (positive or negative) strength $g.$ Its eigenfunctions and spectrum have been found in Refs.~\cites{mcguire_impurity_fermions_65,mcguire_impurity_fermions_66}. The eigenstates were represented as a sum of a product of plane waves with special coefficients. Such a representation is common for the Bethe Ansatz-solvable models. The solution~\cites{mcguire_impurity_fermions_65,mcguire_impurity_fermions_66} may be obtained from the one for the Gaudin-Yang model~\cites{gaudin_fermions_spinful_67,yang_fermions_spinful_67,gaudin_book} as a particular case. However, the model we consider here has its own specificity: there exists a representation for its eigenstates through a single determinant resembling the Slater determinant for the free Fermi gas~\cites{edwards_impurity_90,castella_mob_impurity_93,recher_TGtoFF_det_PainleveV}. We use this representation as a starting point to investigate the time-dependent two-point impurity correlation function. We also argue that this function is the same as the one for a single mobile impurity interacting with the Tonks-Gurardeau gas.

Our main result is an exact expression for the aforementioned correlation function in the limit of infinite system size, $L\to\infty,$ and for arbitrary chemical potential (or density). This representation is obtained for the repulsive as well as attractive impurity-gas $\delta$-function interaction potential of arbitrary strength $g,$ and for arbitrary temperature. The key ingredient is the Fredholm determinant of a linear integral operator of integrable type (see, e.g, section XIV.1 of Ref.~\cite{korepin_book}). The present solution encompasses two particular cases known so far (i) Infinite repulsion $g\to\infty$ \cites{izergin_impenetrable_bosefermi_short_97,izergin_impenetrable_fermions_98} (ii) Arbitrary repulsion $g\ge0$ at zero temperature~\cite{gamayun_impurity_Green_FTG_14}.

The paper is organized as follows: In section~\ref{sec:model} we define the model under consideration. In section~\ref{sec:results} we summarize our main results. In section~\ref{sec:mobrefframe} we explain the transformation from the laboratory to the mobile impurity reference frame. In section \ref{sec:Bethe_Ansatz} we write the eigenfunctions in the mobile impurity reference frame. They are represented as a determinant differing from the Slater determinant representation of the free Fermi gas by a single row only. Using this representation we obtain the form-factors (the overlaps of the eigenfunctions of the model under consideration with the eigenfunctions of the free Fermi gas) in the form of the determinants of some finite-dimensional matrices. In section~\ref{sec:corrrep} we perform a form-factors summation in the case $g\ge 0$ which leads us to the Fredholm determinant representation of the impurity correlation function. In section~\ref{sec:corrat} we extend the results obtained in section~\ref{sec:corrrep} to the case $g < 0.$ Some properties of form-factors and Fredholm determinants, which we use to derive our results, are given in the appendices. 

\section{Model \label{sec:model}}

In this section we define our model. We consider a free one-dimensional spinless Fermi gas at temperature $T$ in the presence of a single distinct quantum particle (impurity) propagating through it. We investigate the correlation function
\begin{equation}\label{Gdef}
G(x,t)= \frac1{Z} \langle \psi_\downarrow(x,t) \psi_\downarrow^\dagger(0,0) \rangle_T.
\end{equation}
The canonical creation (annihilation) operators $\psi_\sigma^\dagger$ ($\psi_\sigma$) carry the subscript $\sigma = \uparrow$ for the Fermi gas, and $\sigma = \downarrow$ for the impurity. The canonical anticommutation relations read
\begin{equation}
\psi_\sigma(x) \psi^\dagger_{\sigma^\prime}(x^\prime)+ \psi^\dagger_{\sigma^\prime}(x^\prime) \psi_\sigma(x) = \delta_{\sigma\sigma^\prime}\delta(x-x^\prime). \label{eq:acomm}
\end{equation}
The temperature-weighted average $\langle \cdots\rangle_T$ in Eq.~\eqref{Gdef} is defined as
\begin{equation}
\langle \cdots \rangle_T = \sum_{\{N\}} \langle 0_\downarrow| \otimes \langle N| e^{-\beta (E_N -\mu N) } \cdots |N\rangle \otimes |0_\downarrow\rangle. \label{eq:thermal}
\end{equation}
Here,
\begin{equation}
|N\rangle = c^\dagger_{p_1 \uparrow} \cdots c^\dagger_{p_N \uparrow} |0_\uparrow \rangle \label{eq:Fermigasstate}
\end{equation}
is the free Fermi gas state containing $N$ fermions with the momenta $p_1, \ldots, p_N.$ The gas is on a ring of circumference $L,$ and periodic boundary conditions are imposed. We have
\begin{equation}
\psi_\sigma^\dagger(x)= \frac1{\sqrt L} \sum_p e^{-ipx} c^\dagger_{p\sigma}, \qquad p=\frac{2\pi n}L, \qquad n=0,\pm1,\pm2,\ldots. \label{eq:psi_to_c}
\end{equation}
The vacuum $|0_\sigma \rangle$ is the state with no particles, $c_\sigma |0_\sigma \rangle=0.$ The sum in Eq.~\eqref{eq:thermal} runs through all possible values of $p_1, \ldots, p_N,$ and $N:$
\begin{equation}
\sum_{\{N\}} = \sum_{N=0}^\infty \sum_{p_1} \cdots \sum_{p_N}.
\end{equation}
The parameter $\beta$ is the inverse temperature, $\beta=1/T.$ The Boltzmann constant, $k_B$, and the Planck constant, $\hbar$, are equal to one in our units. The Boltzmann-Gibbs weight $e^{-\beta (E_N - \mu N) }$ is defined by the value of the free Fermi gas energy $E_N$ in the state~\eqref{eq:Fermigasstate}:
\begin{equation}
H_\uparrow|N\rangle = E_N|N\rangle, \qquad E_N \equiv E_N (p_1,\ldots, p_N) \label{eq:Hff->E}
\end{equation}
and by the chemical potential $\mu$. The Hamiltonian of the free Fermi gas reads
\begin{equation}
H_\uparrow=\int_0^L d x\, \psi_\uparrow^\dagger(x)\left(- \frac{1}{2m}\frac{\partial^2}{\partial x^2} \right)\psi_\uparrow(x),
\label{eq:Hff}
\end{equation}
where $m$ is the particle mass. Therefore
\begin{equation}
E_N = \sum_{j=1}^N \epsilon(p_j), \label{eq:Eff}
\end{equation}
where
\begin{equation}
\epsilon(q) = \frac{q^2}{2m}. \label{eq:dispersion}
\end{equation}
The grand partition function $Z$ in Eq.~\eqref{Gdef} is
\begin{equation}
Z = \sum_{\{N\}} e^{-\beta(E_N - \mu N)} = \prod_p (1+ e^{-\beta[\epsilon(p)-\mu]}). \label{eq:Z}
\end{equation}
Here, the product is taken over all possible values of the single-particle momentum, $p=2\pi n/L,$ $n=0,\pm 1,\pm 2,\ldots.$

The operator $\psi_\downarrow$ in Eq.~\eqref{Gdef} evolves with time $t$ as $\psi_\downarrow(x,t)=  e^{ i t H} \psi_\downarrow(x,0) e^{- i t H},$ where
\begin{equation}\label{Ham}
H= H_\uparrow + H_\mathrm{imp}
\end{equation}
is the Hamiltonian of the entire system, and
\begin{equation}
H_\mathrm{imp}=\int_0^L d x\, \left[ \psi_\downarrow^\dagger(x) \left(- \frac{1}{2m}\frac{\partial^2}{\partial x^2} \right)\psi_\downarrow(x)
+g \psi_\up^\dagger(x)\psi_\downarrow^\dagger(x) \psi_\downarrow(x)\psi_\up(x) \right].
\label{eq:Himp}
\end{equation}
The Hamiltonian~\eqref{Ham} defines the fermionic Gaudin-Yang model~\cites{gaudin_fermions_spinful_67, yang_fermions_spinful_67, gaudin_book}, in which the number of impurity particles, 
\begin{equation}
N_\downarrow = \int_0^L d x\, \psi_\downarrow^\dagger(x) \psi_\downarrow(x),
\end{equation}
is arbitrary. However, only states of the Hamiltonian~\eqref{Ham} with $N_\downarrow=0$, and $N_\downarrow=1$ are relevant for the dynamics described by the correlation function~\eqref{Gdef}. The parameter $g$ in Eq.~\eqref{eq:Himp} gives the strength of the impurity-gas interaction. We consider the case of both repulsive, $g\ge0$, and attractive, $g<0$, interactions. The first-quantized form of the Hamiltonian~\eqref{Ham} with $N_\downarrow=1$ is
\begin{equation}
H = - \frac1{2m} \sum_{j=1}^N\frac{\partial^2}{\partial x_j^2} - \frac1{2m} \frac{\partial^2}{\partial x_\downarrow^2} + g\sum_{j=1}^N \delta(x_j-x_\downarrow). \label{eq:h1}
\end{equation}
Here, $x_1,\ldots,x_N$ are coordinates of the gas particles, and $x_\downarrow$ is the coordinate of the impurity.

Let us now consider the Tonks-Girardeau gas~\cites{tonks_complete_1936,girardeau_impurity_TG_60}. It consists of bosons interacting with each other through a $\delta$-function repulsive potential of infinite strength. The eigenfunctions of the Tonks-Girardeau gas, $\Psi_B(x_1,\ldots,x_N)$, and those of the free Fermi gas, $\Psi_F(x_1,\ldots,x_N)$, are in one-to-one correspondence:
\begin{equation}
\Psi_F(x_1,\ldots,x_N) = \Psi_B(x_1,\ldots,x_N) \prod_{1\le j<l \le N} \sign (x_j-x_l). \label{eq:fb}
\end{equation}
Here, $\sign(x)$ stands for the sign function, equal to one (minus one) for $x$ positive (negative). Adding a single mobile impurity to the Tonks-Girardeau gas leads to the model whose eigenstates $\Psi_B(x_\downarrow, x_1,\ldots,x_N)$ satisfy
\begin{equation}
\Psi_F(x_\downarrow,x_1,\ldots,x_N) = \Psi_B(x_\downarrow,x_1,\ldots,x_N) \prod_{1\le j<l \le N} \sign (x_j-x_l), \label{eq:fi}
\end{equation}
where $\Psi_F(x_\downarrow,x_1,\ldots,x_N)$ is an eigenstate of the Hamiltonian~\eqref{eq:h1}. We stress that $\Psi_B(x_\downarrow,x_1,\ldots,x_N)$ ($\Psi_F(x_\downarrow,x_1,\ldots,x_N)$) is symmetric (antisymmetric) with respect to any permutation of $x_1,\ldots,x_N$. The impurity is distinguishable from the gas particles, therefore no symmetry restriction needs to be enforced when exchanging $x_\downarrow$ with $x_1,\ldots,x_N$. Equations~\eqref{eq:fb} and~\eqref{eq:fi} imply that the correlation function~\eqref{Gdef} is the same whether the impurity is injected into the free Fermi gas or the Tonks-Girardeau gas. Note that this statement holds true for certain other observables (for example, the average impurity momentum~\cite{mathy_flutter_2012}). We present our calculations for an impurity injected into a free Fermi gas, bearing in mind that our results, which are summarized in section~\ref{sec:results}, are applicable to the case of the impurity injected into the Tonks-Girardeau gas as well.

\section{Results \label{sec:results}}

In this section we show the main result of our paper: The Fredholm determinant representation of the correlation function~\eqref{Gdef} at a given temperature $T$ and chemical potential $\mu.$ The structure of the representation crucially depends on the sign of the impurity-gas interaction $g$.

In section~\ref{sec:results_rep} we give the representation for the repulsive impurity-gas interaction, $g \ge 0.$ The result for the attractive interaction, $g< 0,$ is given in section \ref{sec:results_attr}. The particular cases of the infinitely strong repulsion, $g\to\infty$, and attraction, $g\to-\infty$, are discussed in a separate section~\ref{sec:results_infrepattr}.

\subsection{Impurity-gas repulsion, $g \ge 0$ \label{sec:results_rep}}

The Fredholm determinant representation for the correlation function~\eqref{Gdef} in the case of the repulsive impurity-gas interaction, $g\ge0,$ reads
\begin{equation}
G_\mathrm{rep}(x,t) = \frac1{2\pi} \int_{-\infty}^{\infty} d\Lambda\, [(h-1) \det(\hat{I}+\hat{V}) + \det(\hat{I}+\hat{V} - \hat{W})]. \label{eq:Grep}
\end{equation}
Here,
\begin{equation}
h= \frac1{2\pi i} \int_{-\infty}^\infty dq \, e^{-i\tau(q)} [\nu_-(q)-\nu_+(q)] = \frac{1}{\pi} \frac{g^2}4 \int_{-\infty}^\infty dq\,  \frac{e^{-i\tau(q)}}{|k_+ - q|^2}, \label{eq:hxtdef}
\end{equation}
where
\begin{equation}
\tau(q)= t\epsilon(q) - x q, \label{eq:tau1}
\end{equation}
the function $\epsilon(q)$ is defined by Eq.~\eqref{eq:dispersion},
\begin{equation}
\nu_\pm(q) = \frac{g}2 \frac1{q - k_\mp}, \label{eq:nupmdef}
\end{equation}
and
\begin{equation}
k_\pm = \frac{g}2 (\Lambda\pm i). \label{eq:kpmdef}
\end{equation}
The identity operator is denoted by $\hat I$. The kernels of the linear integral operators $\hat{V}$ and $\hat{W}$, on the domain $[-\infty,\infty]\times [-\infty,\infty],$ are defined by
\begin{equation}
V (q,q^\prime) = \frac{e_+(q)e_-(q^\prime)-e_-(q)e_+(q^\prime)}{q-q^\prime}, \label{eq:Vkernel}
\end{equation}
and 
\begin{equation}
W (q,q^\prime) = e_+(q) e_+(q^\prime), \label{eq:Wkernel}
\end{equation}
respectively. The functions $e_\pm$ are defined as
\begin{equation}
e_+(q) = e(q)e_-(q), \qquad e_-(q)= \frac1{\sqrt{\pi}} \sqrt{\vartheta(q)} e^{i\tau(q)/2}, \label{eq:epmdef}
\end{equation}
where
\begin{equation}
e(q) = \frac1{2\pi i} \int_{-\infty}^\infty dq^\prime \, e^{-i\tau(q^\prime)} \left(\frac{\nu_-(q^\prime)}{q^\prime -q-i0} - \frac{\nu_+(q^\prime)}{q^\prime-q+i0}\right), \label{eq:edef}
\end{equation}
and
\begin{equation}
\vartheta(q) = \frac1{e^{\beta [\epsilon(q)-\mu]}+1}
\end{equation}
is the Fermi weight. The determinant -- ``$\det$'' symbol in Eq.~\eqref{eq:Grep} -- stands for the Fredholm determinant of the corresponding linear integral operator. For completeness, we discuss the properties of Fredholm determinants, which we use to derive our results, in~\ref{sec:Fredholm}.

The kernel~\eqref{eq:Vkernel} belongs to a class of integrable kernels~\cites{its_diffeq_corrfunctions_90,korepin_book}. Due to this property the resolvent operator $\hat R$, defined as
\begin{equation}
\hat I - \hat R = (\hat I + \hat V)^{-1},
\end{equation}
is also integrable:
\begin{equation}
R(q,q^\prime) = \frac{f_{+}(q)f_-(q^\prime)-f_{-}(q)f_+(q^\prime)}{q-q^\prime}. \label{eq:R}
\end{equation}
The functions $f_\pm$ are the solutions to the integral equations
\begin{equation}
f_\pm(q) + \int\limits_{-\infty}^\infty dq^\prime \, V(q,q^\prime)f_\pm(q^\prime) = e_\pm(q). \label{eq:f}
\end{equation}
Using the fact that the operator $\hat W$, with the kernel~\eqref{eq:Wkernel}, has rank one, we recast the representation~\eqref{eq:Grep} into the following form:
\begin{equation}
G_\mathrm{rep}(x,t) = \frac{1}{2\pi} \int_{-\infty}^\infty d\Lambda\, (h-B_{++})\det(\hat I + \hat V), \label{eq:Grep2}
\end{equation}
where
\begin{equation}
B_{++} = \int_{-\infty}^\infty dq\, e_+(q)f_+(q).
\end{equation}

\subsection{Impurity-gas attraction, $g<0$ \label{sec:results_attr}}

The Fredholm determinant representation for the correlation function~\eqref{Gdef} in the case of the attractive impurity-gas interaction, $g < 0,$ reads
\begin{equation}
G_\mathrm{attr}(x,t) = G_\mathrm{rep}(x,t) + G^i(x,t). \label{eq:Gattr}
\end{equation}
Here, $G_\mathrm{rep} $ is given by Eq.~\eqref{eq:Grep} in which we let $g$ be negative. The function~$G^i$ in Eq.~\eqref{eq:Gattr} is given by the following expression:
\begin{equation}
\label{GIfinal}
G^i(x,t) = \frac{g^2}{4} \frac1{2\pi}\int_{-\infty}^\infty d\Lambda\, e^{-i[\tau(k_+)+\tau(k_-)]}
[\det(\hat I + \hat V+ \hat V_1)-\det(\hat I + \hat V + \hat V_1 - \hat V_2) ].
\end{equation}
The kernels of the operators entering Eq.~\eqref{GIfinal} are defined on $[-\infty,\infty]\times[-\infty,\infty].$ The kernel of $\hat V$ is given by Eq.~\eqref{eq:Vkernel}, and those of $\hat V_1$ and $\hat V_2$ are given by
\begin{equation}
V_1 (q,q^\prime) = \frac{h e_-(q)e_-(q^\prime)}{(k_+ - q)(k_- - q^\prime)} - \frac{e_-(q) e_+(q^\prime)}{k_- -q} - \frac{e_+(q)e_-(q^\prime)}{k_+ - q^\prime}, \label{eq:V1kernel}
\end{equation}
and
\begin{equation}
V_2 (q,q^\prime) = g^2 \frac{e_- (q) e_-(q^\prime)}{|k_+ - q|^2 |k_+ - q^\prime|^2}, \label{eq:V2kernel}
\end{equation}
respectively. The functions entering Eqs.~\eqref{eq:V1kernel} and~\eqref{eq:V2kernel} are defined in section~\ref{sec:results_rep}. Note that $G^i$ accounts for the bound states formed by the impurity and the gas particles, which exist for the Hamiltonian~\eqref{eq:h1} with $g<0$. This is explicitly shown in section~\ref{sec:corrat}.

\subsection{The limiting cases of infinitely strong repulsion and attraction, $g\to \pm\infty$ \label{sec:results_infrepattr}}

In this section we consider the Fredholm determinant representation for the correlation function~\eqref{Gdef} in the limiting cases of infinitely strong impurity-gas repulsion, $g\to \infty$, and attraction, $g\to -\infty$. In the case $g\to \infty$ our representation coincide with the one given in Refs.~\cites{izergin_impenetrable_bosefermi_short_97,izergin_impenetrable_fermions_98}. In the previously unexplored case $g\to-\infty$ the representation is shown to be identical with the one obtained for $g\to \infty$.

In the $g\to\pm\infty$ limits the function~\eqref{eq:hxtdef} reads as follows:
\begin{equation}
\lim_{g\to\pm\infty} h = 2 \left(\sin\frac{\eta}{2}\right)^2 G_0 (x,t).
\end{equation}
Here, the variable $\eta$ is related to $\Lambda$ by the formula
\begin{equation}
\Lambda = -\cot\left(\frac{\eta}{2}\right),
\end{equation}
and $G_0$ is the Green's function of a free particle,
\begin{equation}
G_0(x,t) = \frac{1}{2\pi} \int_{-\infty}^\infty dk\, e^{-i\tau(k)} = \left\{\begin{array}{ll} e^{-i\frac{\pi}4} \sqrt{\frac{m}{2\pi t}} e^{\frac{im x^2}{2t}} & t\ne 0 \\ \delta(x) & t=0 \end{array} \right. .
\end{equation}
Let us consider the function $G_\mathrm{rep}(x,t)$, which is defined by Eq.~\eqref{eq:Grep} for both repulsive, $g\ge 0$, and attractive, $g<0$, interactions. The $g\to\pm\infty$ limits of this equation read
\begin{equation}
\label{Ginfty}
G_{\infty}(x,t)  
= \frac{1}{2\pi} \int\limits_{-\pi}^{\pi} d\eta\,
%\\ \times 
\left\{[G_0(x,t)-1]\det(\hat I+ \hat V_\infty)+\det(\hat I + \hat V_\infty - \hat W_\infty) \right\}, \qquad g\to \pm\infty .
\end{equation}
The kernels of the operators $\hat V_\infty$ and $\hat W_\infty$ are
\begin{equation}
V_\infty (q,q^\prime) \equiv \lim_{g\to\pm\infty} V(q,q^\prime), \label{eq:Vinfkernel}
\end{equation}
and 
\begin{equation}
W_\infty (q,q^\prime) \equiv \frac{1}{2\left(\sin\frac{\eta}{2}\right)^2}\lim_{g\to\pm\infty} W(q,q^\prime),
\end{equation}
where $V(q,q^\prime)$ and $W(q,q^\prime)$ are defined by Eqs.~\eqref{eq:Vkernel} and~\eqref{eq:Wkernel}, respectively. In the $g\to\pm\infty$ limits the function~\eqref{eq:edef} reads as follows:
\begin{equation}
e_\infty(q) \equiv \lim_{g\to\pm\infty} e(q) = \left(\sin\frac{\eta}{2}\right)^2 \frac{1}\pi \mathrm{p.v.} \int_{-\infty}^\infty dq^\prime \, \frac{e^{-i\tau(q^\prime)}}{q^\prime-q} + \sin\frac{\eta}2 \cos\frac{\eta}{2} e^{-i\tau(q)}, \label{eq:edefgi}
\end{equation}
where the symbol ``p.v.'' indicates that the integral has to be interpreted as the Cauchy principal value. Note that the representation~\eqref{Ginfty} coincides with Eq.~(5.63) of Ref.~\cite{izergin_impenetrable_fermions_98} in the limit $B\to\infty$ and $h\to -\infty$ while keeping $h+B =\mathrm{const}=\mu,$ where $\mu$ is the chemical potential in our model.

The function $G^i(x,t)$ entering Eg.~\eqref{eq:Gattr}, and defined by Eq.~\eqref{GIfinal}, vanishes in the $g\to -\infty$ limit
\begin{equation}
G^i(x,t) = \mathcal{O} \left(\frac{1}g \right), \qquad g\to -\infty. \label{eq:Gi0}
\end{equation}
Let us prove this statement. Letting $g\to-\infty$ in Eqs.~\eqref{eq:Vkernel}, \eqref{eq:V1kernel}, and~\eqref{eq:V2kernel}, after some elementary algebra, we get from~Eq.~\eqref{GIfinal}
\begin{equation}
G^i(x,t) = \frac1{2\pi} \int_{-\infty}^\infty d\Lambda\, e^{-i[\tau(k_+)+\tau(k_-)]} u(\Lambda), \qquad g\to-\infty, \label{eq:Gii2}
\end{equation}
where
\begin{equation}
u(\Lambda) = \frac{4}{(\Lambda^2+1)^2} \det(\hat I +\hat V_\infty) B_{--}.
\end{equation}
Here, the kernel of the operator $\hat V_\infty$ is defined by Eq.~\eqref{eq:Vinfkernel}, and
\begin{equation}
B_{--} = \int_{-\infty}^\infty dq\, e_{-}(q)f_{-}(q),
\end{equation}
where $f_-$ is defined by Eq.~\eqref{eq:f}. We have
\begin{equation}
\tau(k_+)+\tau(k_-) = \frac{t}{m} \frac{g^2}4 (\Lambda^2 -1) -2x\Lambda,
\end{equation}
where $\tau$ is defined by Eq.~\eqref{eq:tau1}, and $k_\pm$ by Eq.~\eqref{eq:kpmdef}. Therefore, the right hand side of Eq.~\eqref{eq:Gii2} decays at least as fast as $1/g$ if
\begin{equation}
|u(0)|<\infty. \label{eq:u}
\end{equation} 
The analytic properties of $u(\Lambda)$  discussed in section 2.3 of Ref.~\cite{cheianov_spin_decoherent_long_04} imply the validity of Eq.~\eqref{eq:u}, thus completing the proof of Eq.~\eqref{eq:Gi0}.

We found that the correlation function~\eqref{Gdef} is given by the same expression~\eqref{Ginfty} for both $g\to\infty$ and $g\to-\infty$ limits. This implies, in particular, that the results for $G_\infty(x,t)$ obtained by the technique developed in Ref.~\cite{zvonarev_ferrobosons_07} (which is specific to the $g\to\infty$ limit) are valid in the $g\to -\infty$ limit.

\section{Mobile impurity reference frame \label{sec:mobrefframe}}

In this section we take our model defined in the laboratory reference frame, and write it in the mobile impurity reference frame, following Ref.~\cite{castella_mob_impurity_93}. We obtain the Hamiltonian which does not contain the impurity coordinate and contains the total momentum (a good quantum number) explicitly. The calculations performed in the rest of our paper are based on this representation of the Hamiltonian.

Let us introduce the operator
\begin{equation}
\mathcal{Q}= e^{iP_\uparrow X_\downarrow}, \label{eq:Qdef}
\end{equation}
where
\begin{equation}
X_\sigma = \int_0^L dx\, x \psi_\sigma^\dagger (x) \psi_\sigma (x), \qquad P_\sigma = \int_0^L dx\, \psi_\sigma^\dagger (x) \left(-i\frac{\partial}{\partial x}\right) \psi_\sigma (x), \qquad \sigma=\uparrow,\downarrow.
\end{equation}
The transformation of an arbitrary operator $\mathcal{O}$ from the laboratory to the mobile impurity reference frame reads
\begin{equation}
\mathcal{O} \to \mathcal{O}_\mathcal{Q} = \mathcal{Q}\mathcal{O}\mathcal{Q}^{-1}. \label{eq:QOQ}
\end{equation}
It is often referred to as a polaron, or the Lee-Low-Pines transformation~\cite{devreese_polaron_09}. For single-particle operators, we get
\begin{align}
&\psi_{\downarrow \mathcal{Q}}(x) = e^{-iP_\uparrow x} \psi_\downarrow(x) \label{eq:psidownQ}\\
&\psi_{\uparrow \mathcal{Q}}(x) = \psi_\uparrow(x-X_\downarrow) ,
\end{align}
and for the Hamiltonian \eqref{Ham} subjected to the condition $N_\downarrow =1$, we get
\begin{equation}\label{eq:Ham_rf}
H_\mathcal{Q}= H_{\uparrow \mathcal{Q}} + H_{\mathrm{imp}\mathcal{Q}}, \qquad N_\downarrow=1.
\end{equation}
Here,
\begin{equation}
H_{\uparrow \mathcal{Q}}= H_{\uparrow }
\label{eq:Hff_rf}
\end{equation}
and
\begin{equation}
H_{\mathrm{imp} \mathcal{Q}}= \frac{(P_\downarrow - P_\uparrow)^2}{2m} + g \rho_\uparrow(0), \qquad N_\downarrow=1.
\label{eq:Himp_rf}
\end{equation}
The total momentum $P$ is conserved in the model~\eqref{Ham}:
\begin{equation}
[P,H]=0, \qquad P=P_\uparrow+ P_\downarrow. \label{eq:PHcomm}
\end{equation}
Equation~\eqref{eq:PHcomm}, written in the mobile impurity reference frame reads
\begin{equation}
[P_\downarrow,H_\mathcal{Q}]=0, \qquad P_\mathcal{Q} = P_\downarrow, \qquad N_\downarrow=1. \label{eq:PHcommref_}
\end{equation}
This means that $P_\downarrow$ is the total momentum of the model written in the mobile impurity reference frame.

The state~\eqref{eq:Fermigasstate} is an eigenfunction of the momentum operator for the free Fermi gas
\begin{equation}
P_\uparrow |N\rangle = P_N|N\rangle, \qquad P_N = \sum_{j=1}^N p_j. \label{eq:pudgs}
\end{equation}
Using that $X_\downarrow |0_\downarrow\rangle =0$ we get for the operator~\eqref{eq:Qdef}
\begin{equation}
\mathcal{Q}|N\rangle \otimes |0_\downarrow\rangle = |N\rangle \otimes |0_\downarrow\rangle. \label{eq:Qgs}
\end{equation}
We note that
\begin{equation}
H|N\rangle \otimes |0_\downarrow\rangle = E_N |N\rangle \otimes |0_\downarrow\rangle,
\label{eq:H|N>|0>}
\end{equation}
where the energy $E_N$ of the free Fermi gas in the state $|N\rangle$ is defined by Eqs.~\eqref{eq:Hff->E} and~\eqref{eq:Eff}. Applying the transformation~\eqref{eq:QOQ} to the operators entering Eq.~\eqref{Gdef} and using Eqs.~\eqref{eq:psidownQ} and~\eqref{eq:pudgs}--\eqref{eq:H|N>|0>} we get
\begin{equation}
G(x,t)= \frac1{Z} \sum_{\{N\}}e^{i E_N t}e^{-iP_N x} e^{-\beta(E_N-\mu N)} \langle 0_\downarrow |\otimes\langle N| \psi_\downarrow(x) e^{-it H_\mathcal{Q}} \psi_\downarrow^\dagger(0) |N\rangle \otimes |0_\downarrow\rangle.
\label{eq:Gdefmobframe}
\end{equation}

Using the momentum space decomposition~\eqref{eq:psi_to_c} for the operators $\psi_\downarrow$ and $\psi^\dagger_\downarrow$ entering Eq.~\eqref{eq:Gdefmobframe}, we get
\begin{equation}
G(x,t) = \frac1{Z} \sum_{\{N\}} e^{-\beta(E_N-\mu N)} e^{i E_N t}  \frac1L \sum_p e^{-i(P_N -p)x} \langle N| e^{-itH_\mathcal{Q}(p)} |N \rangle. \label{eq:Gqt}
\end{equation}
Here, $H_{\mathcal{Q}}(p)$ is obtained from $H_{\mathcal{Q}},$ Eq.~\eqref{eq:Ham_rf}, by projecting the operator $P_\downarrow$ entering Eq.~\eqref{eq:Himp_rf} onto the state $c^\dagger_{p\downarrow} |0_\downarrow\rangle$ having momentum $p:$
\begin{equation}
H_{\mathcal{Q}}(p)=H_\uparrow + H_{\mathrm{imp} \mathcal{Q}}(p), \qquad H_{\mathrm{imp} \mathcal{Q}}(p) = \frac{(p - P_\uparrow)^2}{2m} + g \rho_\uparrow(0). \label{eq:HQq}
\end{equation}

By using the mobile impurity reference frame we reduce the problem of calculating the correlation function~\eqref{Gdef} to the analysis of the matrix elements $\langle N| e^{-itH_\mathcal{Q}(p)} |N \rangle $ entering Eq.~\eqref{eq:Gqt}. The impurity-gas interaction term, $g \rho_\uparrow(0)$, in Eq.~\eqref{eq:HQq} has the form of a static potential scattering off the gas particles. Using the completeness condition
\begin{equation}
\sum_{f_p}|f_p\rangle \langle f_p|=1 \label{eq:fpcomplete}
\end{equation}
for the eigenfunctions $|f_p\rangle$ of the Hamiltonian~$H_{\mathcal Q}(p)$ at a given $p$,
\begin{equation}
H_{\mathcal Q}(p)|f_p\rangle = E_{f}|f_p\rangle, \label{eq:H_Q|f>=E|f>}
\end{equation}
we write
\begin{equation}
\langle N| e^{-itH_\mathcal{Q}(p)} |N \rangle = \sum\limits_{f_p}|\langle N| f_p\rangle|^2 e^{-itE_f}. \label{eq:sum}
\end{equation}
The overlaps $\langle N| f_p\rangle$, often called form-factors, vanish in the large $N$ limit for any $g\ne 0$ (the decay rate for some $\langle N| f_p\rangle$ is found in Ref.~\cite{castella_mob_impurity_93}). This vanishing is an example of the Anderson orthogonality catastrophe~\cite{anderson_infrared_catastrophe_67}. We give an explicit expression for $\langle N| f_p\rangle$ in section~\ref{sec:Bethe_Ansatz}. The sum over $f_p$ in Eq.~\eqref{eq:sum} contains infinitely many terms, and does not vanish as $N\to\infty.$ We calculate this sum in sections~\ref{sec:corrrep} and~\ref{sec:corrat}.

%%%%%%%%%%%%%%%%%%%%%%%%%%%%%%%%%%%%%%%%%%%%%%%%%%%%%%%%%%%%%%%%%%%%%%%%%%%%%%
\section{Bethe Ansatz \label{sec:Bethe_Ansatz}}

In section~\ref{sec:eigenfunctions} we present the eigenfunctions and spectrum of the Hamiltonian~\eqref{eq:HQq}. In section~\ref{sec:Bs} we discuss the solutions of the Bethe Ansatz equations. In section~\ref{sec:overlaps} we calculate the overlaps between the eigenfunctions of the Hamiltonian~\eqref{eq:HQq} and the eigenfunctions~\eqref{eq:Fermigasstate} of the free Fermi gas.

\subsection{Eigenfunctions and spectrum \label{sec:eigenfunctions}}

The eigenfunctions and spectrum of the mobile impurity model~\eqref{eq:h1} were found exactly in Refs.~\cites{mcguire_impurity_fermions_65,mcguire_impurity_fermions_66}. The eigenfunctions in the coordinate representation could be written as a sum running over all permutations of $N$ particles, and containing $N!$ terms. Each of the terms is a product of plane waves multiplied by a factor which does not depend on the coordinates of the particles. Such a structure of the eigenfunctions is common for the Bethe Ansatz solvable models~\cite{korepin_book}. For example, this is the case in the Gaudin-Yang model~\cites{gaudin_fermions_spinful_67,yang_fermions_spinful_67,gaudin_book,takahashi_book}, which contains the mobile impurity model~\eqref{eq:h1} as a particilar instance. However, specific to the model~\eqref{eq:h1}, the eigenfunctions can be written in the mobile impurity reference frame as a single determinant, much resembling the Slater determinant for the eigenfunctions of a free Fermi gas. We demonstrate this explicitly in the present section.

We let the particle mass
\begin{equation}
m=1
\end{equation}
through the rest of the paper in order to lighten notations. We write the eigenfunctions $|f_p\rangle$, defined by Eq.~\eqref{eq:H_Q|f>=E|f>}, in the coordinate representation as the determinant of the $(N+1)\times (N+1)$ matrix
\begin{equation}
\label{f}
f_p(x_1,\dots,x_N) =
\frac{Y_f}{\sqrt{N! L^N}} \begin{vmatrix}
e^{ik_1 x_1} &\dots & e^{ik_{N+1} x_1}\\
\vdots & \ddots & \vdots \\
e^{ik_1 x_N} &\dots & e^{ik_{N+1} x_N}\\
\nu_{-}(k_1) & \dots  & \nu_{-}(k_{N+1})\\
\end{vmatrix}.
\end{equation}
Here, $\nu_-$ is defined by Eq.~\eqref{eq:nupmdef}. The factor $Y_f$ ensures the normalization condition
\begin{equation}
\langle f_p|f^\prime_p \rangle = \delta_{ff^\prime}. \label{eq:fpnorm}
\end{equation}
Note that $\langle f_p|f^\prime_{p^\prime} \rangle \ne \delta_{ff^\prime} $ for $p \ne p^\prime.$ We calculate $Y_f$ in section~\ref{sec:overlaps}. The set of quasi-momenta $k_1,\ldots,k_{N+1}$ satisfies a system of non-linear equations (Bethe equations)
\begin{equation}
\label{Bethe1}
\cot \frac{k_jL}{2} = \frac{2k_j}{g} - \Lambda, \qquad j=1,2,\ldots, N+1,
\end{equation}
and
\begin{equation}
\label{Bethe2}
p = \sum\limits_{j=1}^{N+1}k_j.
\end{equation}
Recall that $p$ is the total momentum of the system in the laboratory reference frame, and it is quantized as follows
\begin{equation}
p= \frac{2\pi}L n, \qquad n=0,\pm1,\pm2,\dots. \label{eq:p}
\end{equation}
The energy $E_f$ of the state $|f_p\rangle$ is
\begin{equation}
\label{energy}
E_f = \frac{1}{2} \sum\limits_{j=1}^{N+1}k_j^2.
\end{equation}
The determinant in Eq.~\eqref{f} differs from the Slater determinant for the eigenfunctions of a free Fermi gas by the last row only. Note that the explicit form of the matrix under the determinant can be changed by a unitary transformation; its size can also be reduced by decomposing the determinant. This explains the variety of the representations equivalent to the one given by Eq.~\eqref{f}, found in the literature~\cites{edwards_impurity_90,castella_mob_impurity_93,recher_TGtoFF_det_PainleveV,mathy_flutter_2012,gamayun_impurity_Green_FTG_14} .

\subsection{Solutions of the Bethe equations \label{sec:Bs}}

In this section we discuss the properties of the solutions of the  Bethe equations~\eqref{Bethe1} and~\eqref{Bethe2} that we use to derive our results.

Any solution $k_1,\ldots,k_{N+1},\Lambda$ of the Bethe equations~\eqref{Bethe1} and~\eqref{Bethe2} has the following properties~\cite{mcguire_impurity_fermions_66}: (i) $\Lambda$ is real. (ii) If $g \ge 0$ all $k_j$'s are real. (iii) If $g<0$ either all $k_j$'s are real, or $k_1,\ldots, k_{N-1}$ are real, while $k_N$ and $k_{N+1}$ have a non-zero imaginary part, and $k_N = k_{N+1}^*$. Furthermore, if $k_N$ and $k_{N+1}$ are complex, they read as follows in the large $L$ limit:
\begin{equation}
\label{roots}
k_N= k_+ + \mathcal{O}(e^{-|g|L}), \qquad k_{N+1} = k_- + \mathcal{O}(e^{-|g|L}),
\end{equation}
where $k_\pm$ are defined by Eq.~\eqref{eq:kpmdef}.

We will often use the following representation of the Bethe equations~\eqref{Bethe1}:
\begin{equation}
e^{i k_j L} = \frac{\nu_{-}(k_j)}{\nu_+(k_j)}, \qquad j=1,\ldots, N+1, \label{eq:Bether}
\end{equation}
where $\nu_{\pm}$ are given by Eq.~\eqref{eq:nupmdef}. Taking the derivative of Eq.~\eqref{eq:Bether} with respect to $\Lambda$ we get
\begin{equation}
\label{deriv}
\frac{\partial k_j}{\partial \Lambda} = \frac{2}{L}\frac{\nu_-(k_j)\nu_+(k_j)}{1+\frac{4}{Lg}\nu_-(k_j)\nu_+(k_j)}, \qquad j=1,\ldots,N+1.
\end{equation}
Substituting Eq.~\eqref{roots} into~\eqref{deriv} we obtain
\begin{equation}
\frac{\partial k_N}{\partial \Lambda} = \frac{\partial k_{N+1}}{\partial \Lambda} = \frac{g}2 + \mathcal{O}(e^{-|g|L}), \qquad k^*_N = k_{N+1}. \label{eq:d}
\end{equation}

Let us introduce the real-valued function $R(x)$ as a solution of the equation
\begin{equation}\label{R}
\cot R(x) = x + \frac{4}{gL} R(x).
\end{equation}
This function is single-valued for $-\infty < x < \infty$ if
\begin{equation}
\frac{4}{gL} \ge -1. \label{eq:Rcond}
\end{equation}
For the rest of the paper we assume that Eq.~\eqref{eq:Rcond} holds true. We note that the  results obtained using this condition are valid for arbitrary $g$ in the $L\to \infty$ limit.

The function $R(x)$ decays monotonously
\begin{equation}
\frac{\partial R(x)}{\partial x} <0, \qquad -\infty < x < \infty \label{eq:dxR}
\end{equation}
and takes the values between $0$ and $\pi$:
\begin{equation}
R(-\infty)=\pi, \qquad R(\infty) =0.
\end{equation}
Using this function, any real-valued quasi-momenta $k_j$ entering Eq.~\eqref{Bethe1} can be parametrized as follows
\begin{equation}\label{eq:kj}
k_j=\frac{2\pi}{L}\left(n_j-\frac{\delta_j}{\pi}\right), \qquad n_j =0,\pm 1,\pm 2,\ldots,
\end{equation}
where
\begin{equation}
\delta_j = R\left(\Lambda- \frac{4\pi}{gL}n_j\right), \qquad 0\le \delta_j < \pi. \label{eq:deltaj}
\end{equation}
Combining Eqs.~\eqref{eq:kj}, \eqref{eq:deltaj}, and~\eqref{eq:dxR} one can immediately see that
\begin{equation}
\frac{\partial k_j}{\partial \Lambda} >0, \qquad -\infty < \Lambda < \infty, \qquad j=1,\ldots,N+1 \label{eq:dk}
\end{equation}
for any real-valued quasi-momentum $k_j.$ 

Evidently, real quasi-momenta $k_j$'s, are in one-to-one correspondence with $n_j$'s. The $k_j$'s  should be different from each other, otherwise the function~\eqref{f} vanishes. This implies that the $n_j$'s should be different as well. Therefore, each eigenstate~\eqref{f} is uniquely characterized by an ordered set $n_1<\cdots<n_{N+1}$ (or by $n_1<\cdots<n_{N-1}$ if $k_N=k_{N+1}^*$) for a given $p.$ We also stress that taking the limits $\Lambda\to\infty$ and $\Lambda\to -\infty$ one gets the same subset of eigenstates~\eqref{f}. The condition $\delta_j < \pi$ imposed in Eq.~\eqref{eq:deltaj} ensures that this subset is taken into account only once.

\subsection{Determinant representation for $\langle N|f_p\rangle$ \label{sec:overlaps}}

In this section we present an explicit formula for the overlaps $\langle N|f_p\rangle$  between the eigenfunctions~\eqref{eq:Fermigasstate} of the free Fermi gas and the eigenfunctions~\eqref{f} of the mobile impurity model. We also present an explicit formula for the normalization constant $Y_f$ in Eq.~\eqref{f}.

Straightforward calculations (see the~\ref{sec:daNf} for details) lead us to the following result:
\begin{equation}
\langle N|f_p\rangle = Y_f\left(\frac{2i}{L}\right)^{N} \det D_f \prod_{j=1}^{N+1} \nu_-(k_j), \label{eq:Nf3}
\end{equation}
where
\begin{equation}
\label{Det1}
\det D_f = \begin{vmatrix}
\frac{1}{k_1-p_1} &\dots & \frac{1}{k_{N+1}-p_1}\\
\vdots & \ddots & \vdots \\
\frac{1}{k_1-p_N} &\dots & \frac{1}{k_{N+1}-p_N}\\
1 & \dots  & 1\\
\end{vmatrix}
\end{equation}
is the determinant of the $(N+1)\times (N+1)$ matrix, and $\nu_-$ is defined by Eq.~\eqref{eq:nupmdef}. We get from Eq.~\eqref{eq:Nf3} 
\begin{equation}
|\langle N|f_p\rangle|^2 = |Y_f|^2\left(\frac{2}{L}\right)^{2N} |\det D_f|^2 \prod_{j=1}^{N+1} |\nu_-(k_j)\nu_+(k_j)|. \label{eq:Nf4}
\end{equation}

Now, using Eq.~\eqref{norma} from the~\ref{sec:aY} we write an explicit expression for $|Y_f|^2$:
\begin{equation}\label{norm}
|Y_f|^{-2} = \prod\limits_{j=1}^{N+1} \left|1+\frac{4}{gL} \nu_{-}(k_j)\nu_{+}(k_j)\right| \left| \sum\limits_{j=1}^{N+1}\frac{\nu_{-}(k_j)\nu_{+}(k_j)} {1+\frac{4}{gL} \nu_{-}(k_j)\nu_{+}(k_j)} \right|.
\end{equation}
Substituting this expression into Eq.~\eqref{eq:Nf4} and using Eq.~\eqref{deriv} we get
\begin{equation}
|\langle N| f_p \rangle|^2 =  \left(\frac{2}{L} \right)^{N} \left|\det D_f\right|^2 \left|\sum\limits_{j=1}^{N+1}\frac{\partial k_j}{\partial \Lambda} \right|^{-1} \left|\prod\limits_{j=1}^{N+1} \frac{\partial k_j}{\partial \Lambda} \right|. \label{over}
\end{equation}
Recall that $k_1,\ldots,k_{N+1},\Lambda$ solve the Bethe equations~\eqref{Bethe1} and~\eqref{Bethe2}.

In sections~\ref{sec:corrrep} and~\ref{sec:corrat} we employ the representation~\eqref{over} to perform the summations in Eqs.~\eqref{eq:sum} and~\eqref{eq:Gqt} and obtain the result expressed in terms of the Fredholm determinants.

\section{Summation in case of impurity-gas repulsion, $g \ge 0$ \label{sec:corrrep}}

In this section we perform the summations in Eqs.~\eqref{eq:sum} and~\eqref{eq:Gqt}. This results in the Fredholm determinant representation for the correlation function~\eqref{Gdef}. We assume that the impurity-gas interaction is repulsive, $g \ge 0.$ The case of the attractive interaction, $g<0,$ is considered in section~\ref{sec:corrat}.

In section~\ref{sec:fpsum} we perform the summation in Eq.~\eqref{eq:sum} using the explicit formula~\eqref{over}. In section~\ref{sec:L} we transform the representations obtained in section~\ref{sec:fpsum}, using the fact that the system size $L\to\infty.$ In section~\ref{sec:N} we perform the summations in Eq.~\eqref{eq:Gqt}.

\subsection{Summation over $f_p$ \label{sec:fpsum}}

In this section we perform the summation over $f_p$ in Eq.~\eqref{eq:sum}.

The quasi-momenta $k_1,\ldots,k_{N+1}$ characterizing the eigenfunctions $f_p$ for a given $p$ are coupled to each other by the condition~\eqref{Bethe2}. There, the total momentum $p$ is quantized, as given by Eq.~\eqref{eq:p}. The quantization is due to a finite system size, $L,$ and should play no role for the observables as $L\to\infty.$ We can therefore write the following identity
\begin{equation}
\delta \left(\sum_{j=1}^{N+1} k_j - p\right) = \frac1{2\pi} \int_{-\infty}^\infty dz \, e^{i\left(\sum_{j=1}^{N+1} k_j -p\right)z}, \qquad L\to\infty. \label{eq:delta}
\end{equation}
The argument $\sum_{j=1}^{N+1}k_j-p$ of the Dirac $\delta$-function on the left hand side of this expression is equal to zero when Eq.~\eqref{Bethe2} is satisfied. Using Eq.~\eqref{eq:delta} to calculate the correlation function~\eqref{Gdef} produces results which are exact in the $L\to\infty$ limit.

Let us now introduce the function
\begin{equation}
\Delta_p(n_1,\ldots,n_{N+1}) = \int_{-\infty}^\infty d\Lambda\, \left| \sum\limits_{j=1}^{N+1}\frac{\partial k_j}{\partial \Lambda} \right| \delta\left(\sum_{j=1}^{N+1}k_j-p\right). \label{eq:D}
\end{equation}
Here, $n_1,\ldots,n_{N+1}$ are integers which define $k_1,\ldots,k_{N+1}$ as given by Eqs.~\eqref{eq:kj} and~\eqref{eq:deltaj}. We stress that $k_j$'s depend on $\Lambda$ through $\delta_j$'s as given by equation~\eqref{eq:deltaj}. Therefore, $\Delta_p(n_1,\ldots,n_{N+1})=1$ if there exists a solution to the Bethe equations \eqref{Bethe1} and \eqref{Bethe2} for a given set $n_1,\ldots,n_{N+1}$, and $p,$ and is equal to zero otherwise:
\begin{equation}
\Delta_p(n_1,\ldots,n_{N+1}) = \left\{\begin{array}{ll}  1 &  \textrm{$n_1,\ldots,n_{N+1}$, and $p$ solve Eqs.~\eqref{Bethe1} and \eqref{Bethe2}} \\ 0 & \textrm{otherwise} \end{array} \right. .
\end{equation}
Defining
\begin{equation}
F_p(n_1,\ldots,n_{N+1}) = \left\{\begin{array}{ll}  |\langle N| f_p \rangle|^2 &  \textrm{$n_1,\ldots,n_{N+1}$, and $p$ solve Eqs.~\eqref{Bethe1} and \eqref{Bethe2}} \\ 0 & \textrm{otherwise} \end{array} \right. ,
\end{equation}
and using Eqs.~\eqref{over} and~\eqref{eq:D} we get
\begin{equation}
F_p(n_1,\ldots,n_{N+1}) = \int_{-\infty}^\infty d\Lambda\, \left(\frac{2}{L} \right)^{N} \left|\prod\limits_{j=1}^{N+1} \frac{\partial k_j}{\partial \Lambda}\right|  \left|\det D_f\right|^2 \delta\left(\sum_{j=1}^{N+1}k_j-p\right). \label{eq:F}
\end{equation}
We stress that $F_p$ is symmetric with respect to the permutations of $n_1,\ldots,n_{N+1}$, and vanishes if any two integers from this set coincide. Using Eq.~\eqref{eq:F} we write the right hand side of Eq.~\eqref{eq:sum} as follows:
\begin{equation}
\langle N| e^{-itH_\mathcal{Q}(p)} |N \rangle = \frac{1}{(N+1)!} \sum_{n_1,\ldots, n_{N+1}} F_p(n_1,\ldots,n_{N+1}) e^{-it E_f}. \label{eq:fs1}
\end{equation}
On the right hand side of this expression the summation over each $n_j$ goes independently over all integers. The factor $1/(N+1)!$ compensates the multiple counting of the Bethe Ansatz solutions stemming from the removal of the ordering condition $n_1<\cdots <n_{N+1}.$ Substituting the identity~\eqref{eq:delta} into Eq.~\eqref{eq:fs1} we get
\begin{multline}
\langle N| e^{-itH_\mathcal{Q}(p)} |N \rangle =\frac{1}{(N+1)!} \int_{-\infty}^\infty d\Lambda\\ 
\times \sum_{n_1,\ldots, n_{N+1}} \int_{-\infty}^\infty \frac{dz}{2\pi} e^{-ip z} \left(\frac{2}{L}\right)^N\left|\det D_f\right|^2\prod\limits_{j=1}^{N+1} \left[e^{-i\tau(k_j)}\left|\frac{\partial k_j}{\partial \Lambda}\right|\right], \label{eq:fs4}
\end{multline}
where
\begin{equation}
\tau(k)= t\frac{k^2}2 - z k. \label{eq:tau}
\end{equation}
We stress that each $k_j$ depends on $n_j$ as given by Eqs.~\eqref{eq:kj} and~\eqref{eq:deltaj}. Since $k_1,\ldots,k_{N+1}$ are real we have
\begin{equation}
\left|\det D_f\right|^2 = \left(\det D_f\right)^2
\end{equation}
and Eq.~\eqref{eq:dk} is applicable. We can, therefore, rewrite Eq.~\eqref{eq:fs4} as follows
\begin{multline}
\langle N| e^{-itH_\mathcal{Q}(p)} |N \rangle =\frac{1}{(N+1)!} \int_{-\infty}^\infty d\Lambda\\ 
\times \sum_{n_1,\ldots, n_{N+1}} \int_{-\infty}^\infty \frac{dz}{2\pi} e^{-ipz} \left(\frac{2}{L}\right)^N\left(\det D_f\right)^2\prod\limits_{j=1}^{N+1} \left[e^{-i\tau(k_j)}\frac{\partial k_j}{\partial \Lambda}\right]. \label{eq:fs2}
\end{multline}

We now transform the summation over $n_1,\ldots, n_{N+1}$ in Eq.~\eqref{eq:fs2} using a technique explained in the \ref{sec:as}. Employing Eq.~\eqref{eq:sn+a3} we get
\begin{equation}
\left(\frac{2}{L}\right)^N \frac{1}{(N+1)!} \sum_{n_1,\ldots, n_{N+1}} \left(\det D_f\right)^2\prod\limits_{j=1}^{N+1} \left[ e^{-i\tau(k_j)}\frac{\partial k_j}{\partial \Lambda} \right] 
 = (h-1) \det  A + \det (A - B). \label{eq:fs3}
\end{equation}
Here,
\begin{equation}
h = \sum_n \frac{\partial k}{\partial \Lambda} e^{-i\tau(k)}, \label{h}
\end{equation}
\begin{equation}
A_{jl} =\frac{2}{L} \sum_n \frac{\partial k}{\partial \Lambda} \frac{e^{-i\tau(k)}}{(k-p_j)(k-p_l)}, \qquad j,l=1,\ldots,N, \label{A}
\end{equation}
and
\begin{equation}
B_{jl}= \frac{2}L e(p_j)e(p_l), \qquad j,l=1,\ldots,N, \label{B2}
\end{equation}
where
\begin{equation}\label{e}
e(q) = \sum_n \frac{\partial k}{\partial \Lambda} \frac{e^{-i\tau(k)}}{k-q}, \qquad q=\frac{2\pi}L n, \qquad n=0,\pm1,\pm2,\ldots .
\end{equation}
We stress that the summation index $n$ in Eqs.~\eqref{h}, \eqref{A}, and \eqref{e} is related to $k$ by Eq.~\eqref{eq:kj}. We rewrite Eq.~\eqref{A} in a form used in our subsequent calculations
\begin{equation}
A_{jl}= \frac{2}L \left[\delta_{jl} \sum_{n} \frac{\partial k}{\partial \Lambda} \frac{e^{-i\tau(k)}}{(k-p_j)^2} + (1-\delta_{jl}) \frac{e(p_j)-e(p_l)}{p_j-p_l} \right], \qquad j,l=1,\ldots,N. \label{A2}
\end{equation}
Substituting Eq.~\eqref{eq:fs3} into~\eqref{eq:fs2} we get
\begin{equation}
\langle N| e^{-itH_\mathcal{Q}(p)} |N \rangle = \int_{-\infty}^\infty d\Lambda  \int_{-\infty}^\infty \frac{dz}{2\pi} e^{-ip z}\left[ (h-1) \det  A + \det (A - B) \right]. \label{G1}
\end{equation}

\subsection{Large $L$ limit \label{sec:L}}

In this section we use the fact that the system size $L\to\infty$ to transform the representations obtained in section~\ref{sec:fpsum}.

We introduce the function
\begin{equation}
b(k) = \frac{2k}g - \Lambda - \cot\frac{kL}{2}
\end{equation}
whose zeroes $b(k)=0$ are the solutions to the Bethe equations~\eqref{Bethe1}. It satisfies the following identity
\begin{equation}
\frac{\partial b(k)}{\partial k} = \left(\frac{\partial k}{\partial \Lambda}\right)^{-1} \qquad \text{at} \qquad b(k)=0.
\end{equation}
Using this identity we transform the summation in Eqs.~\eqref{h}, \eqref{A2}, and \eqref{e} as follows
\begin{equation}\label{ex}
\sum_n \frac{\partial k}{\partial \Lambda} f(k) = \frac{1}{2\pi i} \oint_{\gamma} \frac{dk}{b(k)} f(k).
\end{equation}
The contour $\gamma$ is oriented counter-clockwise. It is chosen in such a way that the zeroes of $b(k)$ are inside the domain bounded by $\gamma$, and the poles of the function $f(k)$ are outside. 

We now take the large $L$ limit. This will remove the effects of the boundary conditions, and further simplify the formulas. The function $\cot(kL/2)$ entering $b(k)$ is periodic with the period $2\pi/L.$ The variation over the period of the other terms entering $b(k)$  can be neglected in the integrals containing $b(k)$ to the leading order of the $1/L$ expansion. Taking into account that
\begin{equation}
\frac{1}{\pi} \int\limits_{0}^{\pi} dk \frac{1}{z-\cot k} = \frac{1}{z+i\mathrm{sign} (\mathrm{Im}\, z)}, \qquad \mathrm{Im}\, z\ne 0
\end{equation}
we can replace $\cot(kL/2)$ with $i$ if $\mathrm{Im}\, \gamma>0$ and with $-i$ if $\mathrm{Im}\, \gamma<0$ in the function $b(k)$ on the right hand side of Eq.~\eqref{ex}.

We begin with applying the arguments from the previous paragraph to the representation~\eqref{ex} of the function~\eqref{h}. We take the contour $\gamma$ consisting of two straight lines, $\gamma^+$ and $\gamma^-$. Here, $\gamma^+$ runs from $\infty + i0$ to $-\infty + i0$, and $\gamma^-$ runs from $-\infty - i0$ to $\infty - i0$. Therefore
\begin{equation}\label{htd}
h = \frac{1}{2\pi i} \int\limits_{-\infty}^{\infty} dk \left(\frac{e^{-i\tau(k)}}{2k/g-\Lambda-i}-\frac{e^{-i\tau(k)}}{2k/g-\Lambda+i}\right).
\end{equation}

We now perform the same procedure with the function $e(q),$ defined by Eq.~\eqref{e}. We take the contour $\gamma$ consisting of $\gamma^+,$ $\gamma^-,$ and $c.$ Here, $c$ is a clockwise-oriented closed contour around the point $q.$ We have
\begin{equation}
\oint_c \frac{dk}{b(k)} \frac1{k-q}=0 \qquad \text{at} \qquad q=\frac{2\pi}L n, \qquad n=0,\pm1,\pm2,\ldots.
\end{equation}
Therefore
\begin{multline}
e(q) = \frac{1}{2\pi i} \int\limits_{-\infty}^{\infty} dk \left(\frac{e^{-i\tau(k)}}{2k/g-\Lambda-i}\frac{1}{k-q-i0}
-\frac{e^{-i\tau(k)}}{2k/g-\Lambda+i}\frac{1}{k-q+i0} \right) \\
= \frac{(2q/g-\Lambda)e^{-i\tau(q)}}{(2q/g-\Lambda)^2+1}+\frac{1}{\pi } \mathrm{p.v.} \int\limits_{-\infty}^{\infty} dk\,
\frac{e^{-i\tau(k)}}{(2k/g-\Lambda)^2+1} \frac{1}{k-q}, \label{eqTD}
\end{multline}
where $\mathrm{p.v.}$ stands for the Cauchy principal value of the integral.

Now we transform the diagonal entries of the matrix~\eqref{A2}:
\begin{equation}
A_{jj}\equiv\frac{2}{L} \sum_n \frac{\partial k}{\partial \Lambda} \frac{e^{-i\tau(k)}}{(k-p_j)^2} =  \frac{1}{2\pi i} \oint\limits_\gamma \frac{dk}{\frac{2k}{g}-\Lambda - \cot \frac{kL}{2}}\frac{e^{-i\tau(k)}}{(k-p_j)^2}.
\end{equation}
We take the same contour $\gamma$ as for $e(q).$ The contour $c$ gives now a non-vanishing contribution. We have
\begin{equation}
A_{jj} = e^{-i\tau(p_j)} + \frac{2}{L}\partial_q e(q)|_{q=p_j}.
\end{equation}
Using
\begin{equation}
\prod_{j=1}^{N} e^{-i\tau(p_j)} = e^{-itE_N +i P_N z},
\end{equation} 
we arrive at the following expressions for the determinants entering~Eq.~\eqref{G1}:
\begin{equation}
\det A = e^{-itE_N +i P_N z} \det (I + \tilde V ), \label{eq:detA}
\end{equation}
and
\begin{equation}
\det (A-B) = e^{-itE_N +i P_N z}  \det (I + \tilde V -\tilde W ). \label{eq:detA-B}
\end{equation}
Here,
\begin{equation}
\tilde V_{j l} = \frac{2\pi}L \frac{E_+(p_j) E_-(p_l) - E_-(p_j)E_+(p_l)}{p_j-p_l}, \qquad j,l=1,\ldots, N \label{eq:Vt}
\end{equation}
and
\begin{equation}
\tilde W_{j l} = \frac{2\pi}L E_+(p_j) E_+(p_l), \qquad j,l=1,\ldots, N. \label{eq:Wt}
\end{equation}
The functions $E_\pm$ are defined as 
\begin{equation}
E_+(q) = e(q) E_-(q), \qquad E_-(q) = \frac1{\sqrt\pi} e^{i\tau(q)/2},
\end{equation}
and the functions $h,$ $e(q)$, and $\tau(q)$ by Eqs.~\eqref{htd}, \eqref{e}, and~\eqref{eq:tau}, respectively. Substituting Eqs.~\eqref{eq:detA} and \eqref{eq:detA-B} into Eq.~\eqref{G1} we get
\begin{multline}
\langle N| e^{-itH_\mathcal{Q}(p)} |N \rangle = \int_{-\infty}^\infty d\Lambda \\
\times \int_{-\infty}^\infty \frac{dz}{2\pi} e^{i(P_N-p) z} e^{-iE_N t} \left[ (h-1) \det(I+\tilde V) + \det (I+\tilde V -\tilde W) \right]. \label{eq:nr}
\end{multline}

\subsection{Summation over $p$ and $\{N\}$  \label{sec:N}}

In this section we perform the summation over $p$ and $\{N\}$ in Eq.~\eqref{eq:Gqt} and obtain the Fredholm determinant representation for the correlation function~\eqref{Gdef}.

We replace the sum over $p$ with an integral,
\begin{equation}
\frac1{L} \sum_{p} \to \frac1{2\pi} \int_{-\infty}^\infty dp,
\end{equation}
while taking the $L\to\infty$ limit of Eq.~\eqref{eq:Gqt}, and get
\begin{equation}
G(x,t) = \frac1{Z} \sum_{\{N\}} e^{-\beta(E_N-\mu N)} e^{i E_N t}  \int_{-\infty}^\infty \frac{dp}{2\pi} e^{-i(P_N -p)x} \langle N| e^{-itH_\mathcal{Q}(p)} |N \rangle. \label{eq:gr1}
\end{equation}
The right hand side of Eq.~\eqref{eq:nr} depends on $p$ through the function $e^{-ipz}$ only. Taking this into account and substituting Eq.~\eqref{eq:nr} into \eqref{eq:gr1} we perform the integration over $p$,
\begin{equation}
\int_{-\infty}^\infty \frac{dp}{2\pi} e^{ip(x-z)} = \delta(x-z),
\end{equation}
and then the integration over $z$, to obtain the following expression
\begin{equation}
G(x,t) =  \frac1{2\pi} \int_{-\infty}^\infty d\Lambda 
\frac1{Z} \sum_{\{N\}} e^{-\beta(E_N-\mu N)} [(h-1)\det(I+\tilde V)+\det(I+ \tilde V - \tilde W)]. \label{eq:gr2}
\end{equation}
Here, $h,$ $\tilde V,$ and $\tilde W$ are given by Eqs.~\eqref{htd}, \eqref{eq:Vt}, and~\eqref{eq:Wt}, respectively.

Equation~\eqref{eq:gr2} is the desired Fredholm determinant representation for the correlation function~\eqref{Gdef}. This representation may be written in the following equivalent form (see the~\ref{sec:Fredholm} for a proof)
\begin{equation}
G(x,t) =  \frac1{2\pi} \int_{-\infty}^\infty d\Lambda\, [(h-1)\det(\hat I+\hat V)+\det(\hat I+ \hat V - \hat W)]. \label{eq:gr3}
\end{equation}
Here, $\hat I$ is the identity operator, and $\hat V$ and $\hat W$ are linear integral operators with the kernels
\begin{equation}
V(q,q^\prime) = \sqrt{\vartheta (q)} \tilde V(q,q^\prime) \sqrt{\vartheta (q^\prime)}, \qquad W(q,q^\prime) = \sqrt{\vartheta (q)} \tilde W(q,q^\prime) \sqrt{\vartheta (q^\prime)},
\end{equation}
where 
\begin{equation}
\vartheta(q) = \frac1{e^{\beta[\epsilon(q)-\mu]}+1}.
\end{equation}
is the Fermi weight, and $\epsilon(q)=q^2/2$ is the single-particle energy. The kernels
\begin{equation}
\tilde V(q,q^\prime) = \frac{E_+(q) E_-(q^\prime) - E_-(q)E_+(q^\prime)}{q-q^\prime}
\end{equation}
and
\begin{equation}
\tilde W(q,q^\prime) = E_+(q) E_+(q^\prime)
\end{equation}
are obtained from Eqs.~\eqref{eq:Vt} and~\eqref{eq:Wt}, respectively.

\section{Summation in case of impurity-gas attraction, $g < 0$ \label{sec:corrat}}

In the present section we adapt the approach of section~\ref{sec:corrrep} to the case of attractive impurity-gas interaction, $g<0$. We make our analysis for $L$ and $g$ satisfying Eq.~\eqref{eq:Rcond}. Our results are applicable for all values of $g$ in the $L\to\infty$ limit.

The quasi-momenta $k_N$ and $k_{N+1}$ could be complex for $g<0$ (this is discussed in section~\ref{sec:eigenfunctions}). We split the sum on the right hand side of Eq.~\eqref{eq:sum} into two parts:
\begin{equation}
\langle N| e^{-itH_\mathcal{Q}(p)} |N \rangle = \sum_{f_p^r}|\langle N| f_p\rangle|^2 e^{-itE_f} + \sum_{f_p^i}|\langle N| f_p\rangle|^2 e^{-itE_f}. \label{eq:sumn}
\end{equation}
The sum over $f_p^r$ runs over the real sets $k_1,\ldots, k_{N+1}$, while the sets with $k_N=k_{N+1}^*$ constitute the sum over $f_p^i.$ The approach of section~\ref{sec:corrrep} can be applied to the sum over $f_p^r$ directly. On substituting this sum into Eq.~\eqref{eq:Gqt} it leads to the same Fredholm determinant representation for $G(x,t)$, as given by Eq.~\eqref{eq:gr3} in the case $g \ge 0$.

Let us now consider the sum over $f_p^i.$ Proceeding as in section~\ref{sec:corrrep} we get the analogue of Eq.~\eqref{eq:fs2}:
\begin{multline}
\sum_{f_p^i} |\langle N| f_p \rangle|^2 e^{-it E_f} = \frac{1}{(N-1)!}
\int_{-\infty}^\infty d\Lambda\, \int_{-\infty}^\infty \frac{dz}{2\pi} e^{-ipz} e^{-i[\tau(k_+) +\tau(k_-)]} \\
\times  \left|\frac{\partial k_N}{\partial\Lambda} \frac{\partial k_{N+1}}{\partial\Lambda}\right| \left(\frac{2}{L}\right)^N \sum_{n_1,\ldots, n_{N-1}} \left|\det D_f\right|^2\prod\limits_{j=1}^{N-1} \left[e^{-i\tau(k_j)}\left|\frac{\partial k_j}{\partial \Lambda}\right| \right].
\label{eq:fs4a}
\end{multline}
Since $k_1,\ldots,k_{N-1}$ are real, and $k_N = k_{N+1}^*$, we have
\begin{equation}
\left|\det D_f \right|^2 = - (\det D_f)^2. \label{eq:ad}
\end{equation}
Using Eq.~\eqref{eq:dk} for $k_1,\ldots,k_{N-1}$, Eq.~\eqref{eq:d} for $k_N$ and $k_{N+1}$, and taking into account Eq.~\eqref{eq:ad},
we transform Eq.~\eqref{eq:fs4a} into
\begin{multline}
\sum_{f_p^i} |\langle N| f_p \rangle|^2 e^{-it E_f} = -\frac{1}{(N-1)!} \frac{g^2}4
\int_{-\infty}^\infty d\Lambda\, \int_{-\infty}^\infty \frac{dz}{2\pi} e^{-ip z} e^{-i[\tau(k_+) +\tau(k_-)]}\\
\times \left(\frac{2}{L}\right)^N \sum_{n_1,\ldots, n_{N-1}} \left(\det D_f\right)^2 \prod\limits_{j=1}^{N-1} \left[ e^{-i\tau(k_j)} \frac{\partial k_j}{\partial \Lambda}\right]. \label{eq:fs2a}
\end{multline}

We now transform the summation over $n_1,\ldots, n_{N-1}$ in Eq.~\eqref{eq:fs2a} using the technique explained in the \ref{sec:as}. Employing Eq.~\eqref{eq:sn-a3} and the asymptotic formula~\eqref{roots} we get
\begin{equation}
\left(\frac{2}{L}\right)^N \frac{1}{(N-1)!} \sum_{n_1,\ldots, n_{N-1}} \left(\det D_f\right)^2\prod\limits_{j=1}^{N-1} \left[ e^{-i\tau(k_j)}\frac{\partial k_j}{\partial \Lambda} \right] 
= - \det  C + \det (C + D). \label{eq:fs3a}
\end{equation}
Here,
\begin{equation}
C_{jl} = \frac2{L}\sum_{n}\frac{\partial k}{\partial\Lambda} \frac{e^{-i\tau(k)}(k_{-}-k)(k_{+}-k)}{(k-p_j)(k-p_l)(k_{-}-p_j)(k_{+}-p_l)}, \qquad j,l=1,\ldots,N, \label{eq:C}
\end{equation}
and
\begin{equation}
D_{jl} = \frac2{L} v(p_j) v(p_l), \qquad j,l=1,\ldots,N,
\end{equation}
where
\begin{equation}
v(q) = \frac{k_{-}-k_+}{(k_{-}-q)(k_{+}-q)}, \qquad q=\frac{2\pi}{L}n, \qquad n=0,\pm1,\pm2,\ldots .
\end{equation}
We stress that the summation index $n$ in Eq.~\eqref{eq:C} is related to $k$ by Eq.~\eqref{eq:kj}. We rewrite Eq.~\eqref{eq:C} in a form used in our subsequent calculations:
\begin{equation}
C_{jl} = A_{jl} - \frac2L \frac{e(p_j)}{k_{+}-p_l} - \frac2L \frac{e(p_l)}{k_{-}-p_j} 
+ \frac2L \frac{h}{(k_{-}-p_j)(k_{+}-p_l)}, \qquad j,l=1,\ldots,N . \label{eq:C2}
\end{equation}
Here, $h$, $A$, and $e$ are defined by Eqs.~\eqref{h}, \eqref{A}, and~\eqref{e}, respectively. Substituting Eq.~\eqref{eq:fs3a} into~\eqref{eq:fs2a} we get
\begin{equation}
\sum_{f_p^i} |\langle N| f_p \rangle|^2 e^{-it E_f}
= \frac{g^2}4 \int_{-\infty}^\infty d\Lambda \int_{-\infty}^\infty \frac{dz}{2\pi} e^{-ip z} e^{-i[\tau(k_+) +\tau(k_-)]} \left[\det  C - \det (C + D) \right]. 
\end{equation}

Further transformations of this sum into the Fredholm determinant representation of the function~\eqref{Gdef} can be performed using steps similar to those performed in sections~\ref{sec:L} and~\ref{sec:N}, and we arrive at the results summarized in section~\ref{sec:results_attr}.

\section{Discussion and outlook}

The main result of the present paper is the Fredholm determinant representation for the time-dependent two-point impurity correlation function~\eqref{Gdef}. We consider a particular Bethe Ansatz solvable model: A mobile impurity interacting with a free Fermi gas (or the Tonks-Girardeau gas~\cites{tonks_complete_1936,girardeau_impurity_TG_60}) through a $\delta$-function potential of arbitrary strength $g.$ We extend the approach of Ref.~\cite{gamayun_impurity_Green_FTG_14} to the case where the impurity-gas interaction can be attractive, $g < 0,$ and the temperature can be finite.

Let us discuss how our results can be used to investigate the mobile impurity dynamics. Various asymptotic formulas for a Fredholm determinant can be obtained by formulating the matrix Riemann-Hilbert problem and solving it asymptotically~\cites{korepin_book,deift_RH_97,kitanine_RH_sine_kernel_09,slavnov_sine_kernel_10,kozlowski_RH_sine_kernel_11}. In model~\eqref{Ham} Fredholm determinant representations are known in the $g\to\infty$ limit~\cites{izergin_impenetrable_bosefermi_short_97,izergin_impenetrable_fermions_98}, and corresponding asymptotic solutions of the matrix Riemann-Hilbert problem have been discussed in Refs.~\cites{gohmann_correlations_impenetrable_98,gohmann_correlations_impenetrable_followup_98,cheianov_spin_decoherent_long_04,cheianov_HubbU_08}. However, these solutions are specific to certain regimes in which the temperature-weighted average is different from the one defined by Eq.~\eqref{eq:thermal} (see Refs.~\cite{gohmann_correlations_gas_phase_99} and~\cite{cheianov_spin_decoherent_short_04} for further description of these regimes). The results obtained in Ref.~\cite{gamayun_impurity_Green_FTG_14} and in the present paper make it possible to formulate the matrix Riemann-Hilbert problem at arbitrary $g,$ in the regime where the temperature-weighted average is defined by Eq.~\eqref{eq:thermal}. The large time and distance asymptotic solution of the matrix Riemann-Hilbert problem should be compared with the predictions of Ref.~\cite{zvonarev_ferrobosons_07} (see Eq.~(6) therein) valid when $g\to\infty$ and the temperature is zero. The Fredholm determinant representation is also promising for investigating the spectral function (defined by the Fourier transform of the correlation function~\eqref{Gdef} from space and time to momentum and frequency variables). All existing approaches based on a mapping between microscopic interacting theories and effective free field theories describe a shape of the spectral function only in some vicinity of a singularity (see \cites{lamacraft_impurity_09,zvonarev_Yang_Gaudin_09} and references therein). In contrast, a numerical evaluation of the Fredholm determinant, combined with asymptotic expansions, has been shown for several models to give precise data for correlation and spectral functions everywhere in space-time and momentum-frequency domains~\cites{cheianov_momentum_crossover_05,cheianov_HubbU_08,zvonarev_BoseHubb_09}. We expect that this approach will give precise data for the impurity spectral function in our model as well. Having such data may help verify the phenomena predicted in Ref.~\cite{kantian_competing_regimes_impurity_14}.

The approach developed in the present paper could be applied to other observables in the model. Of particular interest is the time-dependent average momentum of an impurity particle injected with some initial momentum into a free Fermi gas. This problem has been investigated using the Bethe Ansatz~\cite{mathy_flutter_2012}. However, the form-factors summation (which we do analytically in section~\ref{sec:corrrep} for the impurity correlation function) was done numerically in Ref.~\cite{mathy_flutter_2012}. Later, the same problem was addressed with other techniques~\cites{knap_flutter_signatures_2014,burovski_impurity_momentum_2014,lychkovskiy_perpetual_2014,gamayun_quantum_boltzmann_14,gamayun_kinetic_impurity_TG_14}. The existing results were obtained for particular time scales (for example, short time in the case of simulations using time-dependent density matrix renormalization group) and values of the external parameters (for example, weak impurity-gas coupling in the case of calculations done by diagrammatic methods). We expect that, using the Fredholm determinant representation, the average momentum  of the impurity particle could be precisely calculated at all times and for all values of the external parameters.

Let us now discuss how our results could help further develop the Bethe Ansatz approach. Although the Fredholm determinant representation for correlation functions of the Tonks-Girardeau gas was found~\cites{schultz_TG_63,lenard_TG_64,lenard_TG_66}, the search still continues for other Bethe Ansatz solvable models. 
This search has been particularly successful for those models whose many-body wave functions vanish as any two particles approach each other, either on the lattice or in the continuum. In that case, Fredholm determinant representations can be found using the same techniques as those for a free Fermi gas (for example, Wick's theorem can be employed~\cite{zvonarev_string_09}). In the model we consider here the wave functions do not vanish as the impurity approaches the gas particles, unless $g \to\infty$. Nonetheless, we find that the Fredholm determinant representation for the correlation function, given in Eq.~\eqref{Gdef}, exists for any $g$. Whether its existence is due to some mapping between our model and a non-interacting quantum field theory remains an open problem.

The linear integral operators entering the Fredholm determinant representation given in the present paper are of a special, integrable type (see, e.g, section XIV.1 of Ref.~\cite{korepin_book}). Exploiting the properties of operators of this type, helped represent correlation functions of several models as solutions to non-linear differential equations~\cites{sato_holonomic_II_79,jimbo_painleve_80,its_diffeq_corrfunctions_90}. Our results make it possible to try applying  this approach to the mobile impurity model considered here.

In the case $g\to\infty$, Fredholm determinant representations for the correlation function~\eqref{Gdef} of the free Fermi gas and the Tonks-Girardeau gas on the  lattice are found in Refs.~\cite{izergin_impenetrable_hubbard_98} and~\cite{abarenkova_spin_ladder_01}, respectively. A straightforward modification of the approach given in the present paper would lead to a Fredholm determinant representation for the same models with arbitrary $g$.

\appendix

\section{Fredholm determinants for the free Fermi gas \label{sec:Fredholm}}

In this appendix we recall how a Fredholm determinant is defined and prove an identity used in sections~\ref{sec:corrrep} and~\ref{sec:corrat}.

Take an arbitrary $M\times M$ matrix $V.$ We have
\begin{equation}
\det_{1\le j,l\le M}(\delta_{jl}+V_{jl}) = \sum_{N=0}^M \frac1{N!} \sum_{a_1=1}^M \cdots \sum_{{a_N}=1}^M \det_{1\le j,l \le N} V_{a_j a_l}. \label{eq:dd}
\end{equation}
The limit $M\to\infty$ in this expression defines the Fredholm determinant of the operator $\hat I + \hat V:$
\begin{equation}
\det(\hat I+ \hat V) = \sum_{N=0}^\infty \frac1{N!} \sum_{a_1=1}^\infty \cdots \sum_{{a_N}=1}^\infty \det_{1\le j,l \le N} V_{a_j a_l}. \label{eq:dd2}
\end{equation}
The limit $M\to\infty$ may not exist, and even if does, the expression on the right hand side of Eq.~\eqref{eq:dd2} may diverge for some $V$. We assume, however, that the necessary existence and convergence conditions are fulfilled for the operators encountered in our paper. These operators are linear integral operators, and we therefore use the following rigorous formula to define the Fredhom determinant (see, e.g., \cite{smirnov_book_highermathIV}, vol IV, p.24):
\begin{equation}
\det (\hat I + \hat V) = \sum_{N=0}^\infty \frac1{N!} \int_{-\infty}^\infty dk_1 \cdots \int_{-\infty}^\infty dk_N
\begin{vmatrix}
V (k_1,k_1) &\dots & V(k_1,k_N) \\
\vdots & \ddots &\vdots \\
V (k_N,k_1) &\dots & V (k_N,k_N) \\
\end{vmatrix}. \label{eq:Frdet2}
\end{equation}

Let us now prove an identity which we use in sections~\ref{sec:corrrep} and~\ref{sec:corrat}. We take
\begin{equation}
Z_K = \frac1{Z} \sum_{\{N\}} e^{-\beta(E_N-\mu N)} \det(I+\tilde K). \label{eq:ZK}
\end{equation}
Here,
\begin{equation}
Z = \sum_{\{N\}} e^{-\beta(E_N-\mu N)}
\end{equation}
and $\tilde K$ is an arbitrary $N\times N$ matrix whose entries $\tilde K_{jl} \equiv \tilde K_{p_j p_l}$ are functions of $p_1, \ldots, p_N.$ The energy $E_N$ is the sum of single-particle energies
\begin{equation}
E_N = \sum_{j=1}^N \epsilon(p_j).
\end{equation}
The momenta $p_1, \ldots, p_N$ are quantized
\begin{equation}
p_j = \frac{2\pi}{L}n, \qquad n=0,\pm1,\pm2,\ldots
\end{equation}
and the summation over $\{N\}$ is defined as follows
\begin{equation}
\sum_{\{N\}} = \sum_{N=0}^{\infty} \sum_{p_1} \cdots \sum_{p_N}.
\end{equation}
The parameter $\mu$ is the chemical potential, and $\beta$ is the inverse temperature. We are going to demonstrate that Eq.~\eqref{eq:ZK} can be represented as a Fredholm determinant
\begin{equation}
Z_K = \det(\hat I+\hat K) \label{eq:ZKF}
\end{equation}
for which the operator $\hat K$ has the kernel
\begin{equation}
K(q,q^\prime) = \sqrt{\vartheta(q)} \tilde K(q,q^\prime) \sqrt{\vartheta(q^\prime)}. \label{eq:K}
\end{equation}
Here,
\begin{equation}
\vartheta(q) = \frac1{e^{\beta[\epsilon(q)-\mu]}+1}
\end{equation}
is the Fermi weight, and the kernel $\tilde K(q,q^\prime)$ of the operator $\hat {\tilde K}$ is obtained from the entries of the matrix $\tilde K_{p_j p_l}$ as follows:
\begin{equation}
\tilde K_{p_j p_l} \to \frac{2\pi}L \tilde K(q,q^\prime), \qquad q = p_j, \qquad q^\prime = p_l.
\end{equation}
We stress that the equivalence of the representations~\eqref{eq:ZK} and \eqref{eq:ZKF} is known, see, for example, Ref.~\cite{lenard_TG_66}. We prove it here to make our paper self-contained.

We begin showing the equivalence of Eqs.~\eqref{eq:ZK} and \eqref{eq:ZKF} with taking a determinant from Eq.~\eqref{eq:ZK} and writing it as
\begin{equation}
\det_{1\le j,l\le N }(\delta_{jl}+ \tilde K_{p_j p_l}) =
\det_{1\le j,l\le N} \langle 0|c_{p_j }:e^{\sum_{p,p^\prime} c^\dagger_{p} \tilde K_{p p^\prime} c_{p^\prime}}: c^\dagger_{p_l}|0\rangle 
=\langle N|:e^{\sum_{p,p^\prime} c^\dagger_{p} \tilde K_{p p^\prime} c_{p^\prime }}:|N\rangle .
\label{eq:id1}
\end{equation}
The symbol $:\cdots:$ stands for the normal ordering. The state 
\begin{equation}
|N\rangle = c_{p_1}^\dagger \cdots c_{p_N}^\dagger |0\rangle,
\end{equation}
of a free spinless Fermi gas contains $N$ fermions with the momenta $p_1,\ldots,p_N$, and $|0\rangle $ is the vacuum state containing no fermions. We therefore get for Eq.~\eqref{eq:ZK}
\begin{equation}
Z_K = \frac1{Z} \sum_{\{N\}} \langle N|e^{-\beta(H-\mu N)} :e^{\sum_{p,p^\prime} c^\dagger_{p} \tilde K_{p p^\prime} c_{p^\prime }}:|N\rangle. \label{eq:ZKid1}
\end{equation}
Here, $H$ is the Hamiltonian of a free spinless Fermi gas,
\begin{equation}
H-\mu N = \sum_p [\epsilon(p)-\mu] c^\dagger_p c_p \equiv \sum_{p,p^\prime} c^\dagger_p h_{pp^\prime} c_{p^\prime}. \label{eq:Hff2}
\end{equation}

We now take the identity (see, for example, Ref.~\cite{alexandrov_tau_2013} for a proof)
\begin{equation}
e^{\sum_{p,p^\prime} c^\dagger_p A_{p p^\prime} c_{p^\prime} } = :e^{\sum_{p,p^\prime} c^\dagger_p (e^{A}-I)_{p p^\prime} c_{p^\prime} }:,
\end{equation}
where $I$ is the identity operator, $I_{pp^\prime}=\delta_{pp^\prime}$, and $A_{p p^\prime}$ is an arbitrary function of $p$ and $p^\prime$. Using this identity we bring the exponent of Eq.~\eqref{eq:Hff2} to the normal-ordered form
\begin{equation}
e^{-\beta (H-\mu N)} = :e^{\sum_{p,p^\prime} c^\dagger_p (e^{-\beta h}-I)_{pp^\prime} c_{p^\prime} }:. \label{eq:id3}
\end{equation}
We now substitute Eq.~\eqref{eq:id3} into~\eqref{eq:ZKid1} and use the following identity (see, for example, Ref.~\cite{alexandrov_tau_2013} for a proof) 
\begin{equation}
:e^{\sum_{p,p^\prime} c^\dagger_p A_{p p^\prime} c_{p^\prime}}: :e^{\sum_{p,p^\prime} c^\dagger_p B_{p p^\prime} c_{p^\prime}}: 
\ =\ :e^{\sum_{p,p^\prime} c^\dagger_p [A_{p p^\prime}+B_{p p^\prime} + \sum_{p^{\prime\prime}}A_{p p^{\prime\prime}}B_{p^{\prime\prime} p^\prime}] c_{p^\prime}}:. \label{eq:id2}
\end{equation}
Here, $A_{p p^\prime}$ and $B_{p p^\prime}$ are arbitrary functions of $p$ and $p^\prime$. We get
\begin{equation}
Z_K = \frac1Z \sum_{\{N\}} \langle N|: e^{\sum_{p,p^\prime}c^\dagger_{p} [e^{-\beta h}(I + \tilde K)-I]_{pp^\prime} c_{p^\prime}} :|N\rangle. \label{eq:ZK1}
\end{equation}
The sum over $\{N\}$ is the trace over all fermionic states. Therefore Eq.~\eqref{eq:ZK1} can be written as
\begin{equation}
Z_K = \frac{\Tr : e^{\sum_{p,p^\prime}c^\dagger_{p} (e^{-\beta \tilde h}-I)_{pp^\prime} c_{p^\prime}} : }{\Tr : e^{\sum_{p,p^\prime}c^\dagger_{p} (e^{-\beta h}-I)_{pp^\prime} c_{p^\prime}} :}, \label{eq:ZK2}
\end{equation}
where $\tilde h$ is defined by the formula
\begin{equation}
e^{-\beta \tilde h} = e^{-\beta h}( I + {\tilde K}).
\end{equation}
We recognize the numerator and denominator on the right hand side of Eq.~\eqref{eq:ZK2} as the partition functions of the corresponding fermionic models, rewrite them using a standard formula found in statistical mechanics textbooks, and get
\begin{equation}
Z_K = \frac{\det(\hat I +e^{-\beta \hat{\tilde h}})}{\det(\hat I +e^{-\beta  \hat h})} = \frac{\det(\hat I +e^{-\beta \hat h}(\hat I + \hat{\tilde K} ))}{\det(\hat I +e^{-\beta \hat h})} = \det(\hat I +\hat{ K}),
\end{equation}
where $\hat {\tilde K}$ is related to $\hat K$ by Eq.~\eqref{eq:K}. We thus proved that Eqs.~\eqref{eq:ZKF} and~\eqref{eq:ZK} are equivalent. 

\section{Summation formulas at finite $N$}

In this appendix we explain how we insert the summations (or integrations) into the product of two determinants. We use the following definition of the determinant of an arbitrary $N\times N$ matrix $A:$
\begin{equation}
\det A = \sum_{a_1, \ldots, a_N=1}^N \epsilon_{a_1\ldots a_N} A_{1 a_1} \cdots A_{N a_N}. \label{eq:dete}
\end{equation}
Here, $\epsilon_{a_1\ldots a_N} $ is the Levi-Civita symbol
\begin{equation}
\epsilon_{a_1\ldots a_N} = \left\{\begin{array}{ll} +1 & \textrm{if } a_1,\ldots, a_N \textrm{ is an even permutation of } 1,\ldots,N \\ -1 & \textrm{if } a_1,\ldots, a_N \textrm{ is an odd permutation of } 1,\ldots,N \\ 0 & \textrm{otherwise} \end{array} \right. .
\end{equation}
The product of two Levi-Civita symbols with $a_1,\ldots, a_N$ and $b_1,\ldots, b_N$ taking values from the set $1,\ldots,N$ can be written as the determinant of the $N\times N$ matrix whose entries are the Kronecker delta symbols:
\begin{equation}
\epsilon_{a_1\ldots a_N} \epsilon_{b_1\ldots b_N} = \begin{vmatrix}
\delta_{a_1 b_1} &\dots & \delta_{a_1 b_{N}}\\
\vdots & \ddots & \vdots \\
\delta_{a_{N} b_1} &\dots & \delta_{a_{N} b_{N}}\\ 
\end{vmatrix}. \label{eq:detee}
\end{equation}

\subsection{Summation in the overlap $\langle N|f_p\rangle$ \label{sec:daNf}}

We write Eq.~\eqref{f} as
\begin{equation}
|f_p\rangle = \frac{Y_f}{\sqrt{N!L^N}} \prod_{j=1}^{N+1} \nu_-(k_j) \sum_{a_1,\ldots,a_N=1}^N \epsilon_{a_1 \ldots a_N} \prod_{j=1}^N \left[\frac{e^{ik_{a_j}x_j}}{\nu_-(k_{a_j})} - \frac{e^{ik_{N+1}x_j}} {\nu_-(k_{N+1})} \right] \label{eq:fa1}
\end{equation}
and the coordinate representation of state~\eqref{eq:Fermigasstate} as
\begin{equation}
|N\rangle = \frac1{\sqrt{N! L^N}}\sum_{a_1,\ldots,a_N=1}^N \epsilon_{a_1\ldots a_N} \prod_{j=1}^N e^{ip_{a_j}x_j}. \label{eq:na1}
\end{equation}
Recall that the function $\nu_-$ is defined by Eq.~\eqref{eq:nupmdef}. Taking the product of Eq.~\eqref{eq:fa1} with the complex conjugate of Eq.~\eqref{eq:na1}, integrating the result over $x_1,\ldots, x_N$, and using identity~\eqref{eq:detee} we get
\begin{multline}
\langle N|f_p\rangle = \frac{Y_f}{N!}\prod_{j=1}^{N+1} \nu_-(k_j) \sum_{a_1,\ldots,a_N=1}^N \sum_{b_1,\ldots,b_N=1}^N \epsilon_{a_1 \ldots a_N}\epsilon_{b_1 \ldots b_N}\\
\times \prod_{j=1}^N \left[\frac{\phi(k_{a_j},p_{b_j})}{\nu_-(k_{a_j})} - \frac{\phi(k_{N+1},p_{b_j})} {\nu_-(k_{N+1})} \right] = Y_f
\begin{vmatrix}
\phi(k_1,p_1) &\dots & \phi(k_{N+1},p_1) \\
\vdots & \ddots & \vdots \\
\phi(k_1,p_N) &\dots & \phi(k_{N+1},p_N)  \\
\nu_-(k_1) & \dots  & \nu_-(k_{N+1})
\end{vmatrix}. \label{eq:Nfa1}
\end{multline}
Here, the function $\phi$ is defined as
\begin{equation}
\phi(z,z^\prime) = \frac1L \int_0^L dx\, e^{ix(z-z^\prime)}. \label{eq:phia}
\end{equation}

Using the quantization condition for the free Fermi gas momenta, $\exp\{iL p_j\}=1,$ $j=1,\ldots,N$, and Eqs.~\eqref{eq:Bether} and~\eqref{eq:nupmdef} we arrive at
\begin{equation}
\phi(k_a,p_b) = \frac{2i}L \frac{\nu_-(k_a)}{k_a-p_b}, \qquad a=1,\ldots,N+1, \qquad b=1,\ldots,N. \label{eq:varphi1}
\end{equation}
Substituting Eq.~\eqref{eq:varphi1} into~\eqref{eq:Nfa1} we have
\begin{equation}
\langle N|f_p\rangle = Y_f\left(\frac{2i}{L}\right)^{N} \det D_f \prod_{j=1}^{N+1} \nu_-(k_j), \label{eq:Nfa2}
\end{equation}
where
\begin{equation}
\label{Det1app}
\det D_f = \begin{vmatrix}
\frac{1}{k_1-p_1} &\dots & \frac{1}{k_{N+1}-p_1}\\
\vdots & \ddots & \vdots \\
\frac{1}{k_1-p_N} &\dots & \frac{1}{k_{N+1}-p_N}\\
1 & \dots  & 1\\
\end{vmatrix}
\end{equation}
is the determinant of the $(N+1)\times (N+1)$ matrix.

\subsection{Summation in the normalization factor $|Y_f^2|$ \label{sec:aY}}

We write Eq.~\eqref{f} as
\begin{equation}
|f_p\rangle = \frac{Y_f}{\sqrt{N!L^N}} \sum_{a_1,\ldots,a_{N+1}=1}^{N+1}  \epsilon_{a_1\ldots a_{N+1}} \nu_-(k_{a_{N+1}}) \prod_{j=1}^N e^{i k_{a_j} x_j} . \label{eq:dete1}
\end{equation}
Recall that the function $\nu_-$ is defined by Eq.~\eqref{eq:nupmdef}. Taking the product of Eq.~\eqref{eq:dete1} with its complex conjugate, integrating the result over $x_1,\ldots, x_N$, and using identity~\eqref{eq:detee} we get
\begin{multline}
|Y_f|^{-2}= \frac1{N!} \sum_{a_1,\ldots,a_{N+1}=1}^{N+1} \sum_{b_1,\ldots,b_{N+1}=1}^{N+1} \epsilon_{a_1\ldots a_{N+1}} \epsilon_{b_1\ldots b_{N+1}}
\nu_-(k_{a_{N+1}}) \nu_+(k^*_{b_{N+1}}) \prod_{j=1}^N \phi(k_{a_j},k^*_{b_j})
\\
=- \begin{vmatrix}
\phi(k_1,k_1^*) &\dots & \phi(k_{N+1},k^*_1) & \nu_{+}(k^*_1) \\
\vdots & \ddots & \vdots & \vdots \\
\phi(k_1,k_{N+1}^*) &\dots & \phi(k_{N+1},k^*_{N+1}) & \nu_{+}(k^*_{N+1}) \\
\nu_{-}(k_1) & \dots  & \nu_{-}(k_{N+1}) & 0 \\
\end{vmatrix}. \label{eq:Yfdet}
\end{multline}
Here, the function $\phi$ is given by Eq.~\eqref{eq:phia}.

Let us now use the fact that $k_1,\ldots, k_{N+1}$ are solutions to the Bethe equations~\eqref{Bethe1} and~\eqref{Bethe2}. We established in section~\ref{sec:overlaps} that $k_1,\ldots, k_{N-1}$ are always real, and $k_N$ and $k_{N+1}$ are either both real or complex-conjugated to each other, $k_N^* = k_{N+1}$. We thus eliminate the complex conjugation from Eq.~\eqref{eq:Yfdet}:
\begin{equation}
|Y_f|^{-2} =-\alpha \begin{vmatrix}
\phi(k_1,k_1) &\dots & \phi(k_{N+1},k_1) & \nu_{+}(k_1) \\
\vdots & \ddots & \vdots & \vdots \\
\phi(k_1,k_{N+1}) &\dots & \phi(k_{N+1},k_{N+1}) & \nu_{+}(k_{N+1}) \\
\nu_{-}(k_1) & \dots  & \nu_{-}(k_{N+1}) & 0 \\
\end{vmatrix}. \label{eq:Yfdet2}
\end{equation}
Here,
\begin{equation}
\alpha = \left\{\begin{array}{ll}  1 &  \mathrm{Im}\, k_N =0 \\ -1 & \mathrm{Im}\, k_N \ne 0 \end{array} \right. . \label{eq:alphadef}
\end{equation}

Using Eqs.~\eqref{eq:Bether} and~\eqref{eq:nupmdef} we get from Eq.~\eqref{eq:phia} the expression
\begin{equation}
\phi(k_a,k_b) = \delta_{ab} -(1-\delta_{ab})\frac{4}{Lg} \nu_{-}(k_a) \nu_+(k_b), \qquad a,b=1,\ldots,N+1 \label{eq:identity}
\end{equation}
valid for any quasi-momenta $k_a$ and $k_b$ from a given solution $k_1,\ldots,k_{N+1},\Lambda$ of the Bethe equations~\eqref{Bethe1} and~\eqref{Bethe2}. Substituting Eq.~\eqref{eq:identity} into Eq.~\eqref{eq:Yfdet2} we get after some elementary algebra
\begin{equation}\label{norma}
|Y_f|^{-2} = \alpha\prod\limits_{j=1}^{N+1} \left[1+\frac{4}{gL} \nu_{-}(k_j)\nu_{+}(k_j)\right] \sum\limits_{j=1}^{N+1}\frac{\nu_{-}(k_j)\nu_{+}(k_j)} {1+\frac{4}{gL} \nu_{-}(k_j)\nu_{+}(k_j)}.
\end{equation}

\subsection{Two summation formulas with $(\det D_f)^2$ \label{sec:as}}

We consider
\begin{equation}
S = \frac1{(N+1)!}\sum_{n_1,\ldots,n_{N+1}} (\det D_f)^2 \prod_{j=1}^{N+1} f(k_j), \label{eq:sn+a1}
\end{equation}
where $\det D_f$ is defined by Eq.~\eqref{Det1}, and $k_j$'s and $n_j$'s by Eq.~\eqref{eq:kj}. The function $f(k_j)$ is arbitrary. We write Eq.~\eqref{Det1} as
\begin{equation}
\det D_f = \sum_{a_1,\ldots, a_{N+1}=1}^{N+1} \epsilon_{a_1\ldots a_{N+1}} \prod_{j=1}^N  \frac1{k_{a_j}-p_j}.
\end{equation}
Substituting this representation into Eq.~\eqref{eq:sn+a1} and using identity~\eqref{eq:detee} we get after some elementary algebra
\begin{equation}
S = \begin{vmatrix}
\tilde A_{11} &\dots & \tilde A_{1N} & e_1 \\
\vdots & \ddots & \vdots & \vdots \\
\tilde A_{N1} &\dots & \tilde A_{NN} & e_N \\
e_1 & \dots  & e_N & h \\
\end{vmatrix}. \label{eq:sn+a2}
\end{equation}
Here,
\begin{equation}
\tilde A_{jl} = \sum_n \frac{f(k)}{(k-p_j)(k-p_l)}, \qquad j,l=1,\ldots,N,
\end{equation}
\begin{equation}
\qquad e_j = \sum_n \frac{f(k)}{k-p_j}, \qquad j=1,\ldots,N,
\end{equation}
and
\begin{equation}
h = \sum_n f(k).
\end{equation}
We finally write Eq.~\eqref{eq:sn+a2} as
\begin{equation}
S = (h-1)\det_{1\le j,l\le N} \tilde A_{jl} + \det_{1\le j,l\le N} (\tilde A_{jl} -e_j e_l). \label{eq:sn+a3}
\end{equation}

We now consider 
\begin{equation}
s = \frac1{(N-1)!}\sum_{n_1,\ldots,n_{N-1}} (\det D_f)^2 \prod_{j=1}^{N-1} f(k_j), \label{eq:sn-a1}
\end{equation}
where $\det D_f$ is defined by Eq.~\eqref{Det1}. We stress that Eq.~\eqref{eq:sn-a1} contains the parameters $k_N$ and $k_{N+1}$ without a summation over them, in contrast to Eq.~\eqref{eq:sn+a1}. We use the following two representations for Eq.~\eqref{Det1}:
\begin{equation}
\det D_f = \sum_{a_1,\ldots,a_N=1}^N \epsilon_{a_1 \ldots a_N} \frac{k_{N+1}-k_N}{(k_N - p_{a_N})(k_{N+1}-p_{a_N})} \prod_{j=1}^{N-1} \frac{k_{N+1}-k_j}{(k_j-p_{a_j})(k_{N+1}-p_{a_j})},
\end{equation}
and
\begin{equation}
\det D_f = \sum_{a_1,\ldots,a_N=1}^N \epsilon_{a_1 \ldots a_N} \frac{k_{N+1}-k_N}{(k_N - p_{a_N})(k_{N+1}-p_{a_N})} \prod_{j=1}^{N-1} \frac{k_{N}-k_j}{(k_j-p_{a_j})(k_{N}-p_{a_j})}.
\end{equation}
Substituting the product of these representations into Eq.~\eqref{eq:sn-a1} and using the identity~\eqref{eq:detee} we get
\begin{equation}
s = -\begin{vmatrix}
\tilde C_{11} &\dots & \tilde C_{1N} & v_1 \\
\vdots & \ddots & \vdots & \vdots \\
\tilde C_{N1} &\dots & \tilde C_{NN} & v_N \\
v_1 & \dots  & v_N & 0 \\
\end{vmatrix}. \label{eq:sn-a2}
\end{equation}
Here,
\begin{equation}
\tilde C_{jl} = \sum_{n} \frac{f(k)(k_{N+1}-k)(k_N-k)}{(k-p_j)(k-p_l)(k_{N+1}-p_j)(k_N-p_l)}, \qquad j,l=1,\ldots,N,
\end{equation}
and
\begin{equation}
v_j = \frac{k_{N+1}-k_N}{(k_{N+1}-p_j)(k_N-p_j)}, \qquad j=1,\ldots,N.
\end{equation}
We finally write Eq.~\eqref{eq:sn-a2} as
\begin{equation}
s = -\det_{1\le j,l\le N} \tilde C_{jl} + \det_{1\le j,l\le N} (\tilde C_{jl} +v_j v_l). \label{eq:sn-a3}
\end{equation}

%%%%%%%%%%%%%%%%%%%%%%%%%%%%%%%%%%%%%%%%%%%%%%%%%%%%%%%%%%%%%%%%%%%%%%%%%%%%%%
\section*{Acknowledgments}

We thank A.~Maitra, S.~Majumdar, G.~Schehr, and A.~Sykes for careful reading of the manuscript. The work of O.G. and M.B.Z. is part of the Delta ITP consortium, a program of the Netherlands Organisation for Scientific Research (NWO) that is funded by the Dutch Ministry of Education, Culture and Science (OCW). The work of A.G.P. is partially supported by the Russian Science Foundation, Grant No. 14-11-00598.

%%%%%%%%%%%%%%%%%%%%%%%%%%%%%%%%%%%%%%%%%%%%%%%%%%%%%%%%%%%%%%%%%%%%%%%%%%%%%%

%\section*{References}

%\bibliographystyle{apsrmp}
\bibliography{FredholmT-bib}

\end{document}